%% file: main.tex
\documentclass[11pt,a4]{article}
\usepackage{geometry}%[top=0.85in,left=2.75in,footskip=0.75in]

\usepackage{amsmath,amssymb}

\usepackage{changepage}

\usepackage{textcomp,marvosym}

\usepackage{cite}

\usepackage{color}

\usepackage{nameref,hyperref}

\usepackage{makecell}

\usepackage[table]{xcolor}

\usepackage{todonotes}
\usepackage{booktabs}
\usepackage{siunitx}
\usepackage[aboveskip=1pt,labelfont=bf,justification=raggedright]{caption}
\usepackage{booktabs}
\usepackage{nameref}
\usepackage{longtable}
\usepackage[longtable]{multirow}
\usepackage{afterpage}  % load the afterpage package

\input{macros.tex}

\newcommand{\TODO}[1]{}
\renewcommand{\TODO}[1]{{\color{red} TODO: {#1}}}

\usepackage{array}

\newcolumntype{+}{!{\vrule width 2pt}}

\newlength\savedwidth

\usepackage{caption} %[aboveskip=1pt,labelfont=bf,labelsep=period,justification=raggedright,singlelinecheck=off]

\makeatletter
\renewcommand{\@biblabel}[1]{\quad#1.}
\makeatother

\begin{document}

\begin{flushleft}
{\Large
\textbf\newline{Mapping the DeFi Crime Landscape: An Evidence-based Picture} % Please use "sentence case" for title and headings (capitalize only the first word in a title (or heading), the first word in a subtitle (or subheading), and any proper nouns).
}
\newline
\\
Catherine Carpentier-Desjardins \textsuperscript{1},
Masarah Paquet-Clouston \textsuperscript{1,2,*},
Stefan Kitzler \textsuperscript{2,3} and
Bernhard Haslhofer \textsuperscript{2}
 %with the Lorem Ipsum Consortium\textsuperscript{\textpilcrow}
\\
\bigskip
\textbf{1} School of Criminology, Université de Montréal, Montréal, Canada
\\
\textbf{2} Complexity Science Hub Vienna, Vienna, Austria
\\
\textbf{3} AIT Austrian Institute of Technology, Vienna, Austria
\\
\textbf{*} Corresponding author: School of Criminology, Université de Montréal, Pav. Lionel-Groulx
3150, Jean-Brillant Street,
Montréal (QC)
H3T 1N8. m.paquet-clouston@umontreal.ca
\\
\bigskip
The full paper is now published in the \textit{Journal of Cybersecurity} available at: https://academic.oup.com/cybersecurity/article/11/1/tyae029/7962044

\end{flushleft}

\input{sections/abstract}

%\linenumbers

\input{sections/introduction}

\input{sections/literature}

\input{sections/data_methods}

\input{sections/results}

\input{sections/discussion}

\input{sections/conclusion}

\section*{Acknowledgments}
We would like to thank Liuhuaying Yang for her work on the infographics and her patience working with us. We would also like to thank the administrators of De.Fi REKT, SlowMist and Cryptosec for maintaining the crime event databases; and specifically, the admins of De.Fi REKT for answering our questions and providing us with a research API access. This research was funded with an insight development grant from the Social Sciences and Humanities Research Council (SSHRC) as well as the Human Centric Cybersecurity Partnership (HC2P). It was also partially funded by the Austrian security research program KIRAS of the Federal Ministry of Finance (BMF) under the project DeFiTrace (grant agreement 905300), the FFG BRIDGE project AMALFI (grant agreement 898883), and the COMET Centre ABC (Austrian Blockchain Center) managed by the FFG (grant agreement 909237).

\bibliography{finalreferences_3}

\clearpage

\input{sections/annex}

\end{document}

%% file: macros.tex
\def\NallDb/{\num{1141}}
\def\NallDbTwoThousandSeventeen/{\num{16}}
\def\NallDbTwoThousandEighteen/{\num{68}}
\def\NallDbTwoThousandNineteen/{\num{103}}
\def\NallDbTwoThousandTwenty/{\num{224}}
\def\NallDbTwoThousandTwentyOne/{\num{307}}
\def\NallDbTwoThousandTwentyTwo/{\num{435}}
\def\NMinFinDam/{\$158}
\def\NMeanFinDam/{\$29.6M}
\def\NStdFinDam/{\$205.9M}
\def\NMedianFinDam/{\$353.8k}
\def\NMaxFinDam/{\$3.6B}
\def\NSumFinDam/{\$29.5B}
\def\NallDbHasFinDam/{\num{984}}
\def\NCeFi/{\num{105}}
\def\NCeFiHasFinDam/{\num{80}}
\def\NNonCeFi/{\num{1036}}
\def\NNonCeFiHasFinDam/{\num{904}}
\def\NDeFi/{\num{1036}}
\def\NDeFiDismissed/{\num{0}}
\def\NCeFiEmbezzlement/{\$1.1B}
\def\NCeFiPonzi/{\$7.3B}
\def\PActorTarget/{\num{52}}
\def\PActorTargetMainTechnicalVulnerability/{\num{46.9}}
\def\PActorTargetMainTechnicalVulnerabilityGeneralContractVulnerability/{\num{30.3}}
\def\PActorTargetMainTechnicalVulnerabilityGeneralContractVulnerabilitySpecificLogicalBugCustomFlaw/{\num{5.2}}
\def\PActorTargetMainTechnicalVulnerabilityGeneralContractVulnerabilitySpecificRollbackAttack/{\num{1.7}}
\def\PActorTargetMainTechnicalVulnerabilityGeneralContractVulnerabilitySpecificRandomNumberAttack/{\num{1.8}}
\def\PActorTargetMainTechnicalVulnerabilityGeneralContractVulnerabilitySpecificUndetermined/{\num{5.6}}
\def\PActorTargetMainTechnicalVulnerabilityGeneralContractVulnerabilitySpecificAccessControlFlaw/{\num{12.1}}
\def\PActorTargetMainTechnicalVulnerabilityGeneralContractVulnerabilitySpecificKValueVerificationVulnerability/{\num{0.4}}
\def\PActorTargetMainTechnicalVulnerabilityGeneralContractVulnerabilitySpecificReentrancy/{\num{3.1}}
\def\PActorTargetMainTechnicalVulnerabilityGeneralContractVulnerabilitySpecificIntegerOverflow/{\num{0.4}}
\def\PActorTargetMainTechnicalVulnerabilityGeneralTransactionAttack/{\num{2.4}}
\def\PActorTargetMainTechnicalVulnerabilityGeneralTransactionAttackSpecificTransactionCongestionAttack/{\num{2}}
\def\PActorTargetMainTechnicalVulnerabilityGeneralTransactionAttackSpecificReplayAttack/{\num{0.3}}
\def\PActorTargetMainTechnicalVulnerabilityGeneralTransactionAttackSpecificFrontRunningAttack/{\num{0.1}}
\def\PActorTargetMainTechnicalVulnerabilityGeneralHackedExploitedInfrastructure/{\num{4.5}}
\def\PActorTargetMainTechnicalVulnerabilityGeneralHackedExploitedInfrastructureSpecificAccessingPrivateKeysData/{\num{3.8}}
\def\PActorTargetMainTechnicalVulnerabilityGeneralHackedExploitedInfrastructureSpecificBGPHijacking/{\num{0.1}}
\def\PActorTargetMainTechnicalVulnerabilityGeneralHackedExploitedInfrastructureSpecificRansomware/{\num{0.4}}
\def\PActorTargetMainTechnicalVulnerabilityGeneralHackedExploitedInfrastructureSpecificUndetermined/{\num{0.2}}
\def\PActorTargetMainTechnicalVulnerabilityGeneralInterconnectedActorsFlaw/{\num{7}}
\def\PActorTargetMainTechnicalVulnerabilityGeneralInterconnectedActorsFlawSpecificUndetermined/{\num{0.3}}
\def\PActorTargetMainTechnicalVulnerabilityGeneralInterconnectedActorsFlawSpecificArbitragewithFlashLoan/{\num{3.5}}
\def\PActorTargetMainTechnicalVulnerabilityGeneralInterconnectedActorsFlawSpecificOracleManipulation/{\num{3.1}}
\def\PActorTargetMainTechnicalVulnerabilityGeneralDecentralizationIssue/{\num{2}}
\def\PActorTargetMainTechnicalVulnerabilityGeneralDecentralizationIssueSpecificAttack/{\num{1.8}}
\def\PActorTargetMainTechnicalVulnerabilityGeneralDecentralizationIssueSpecificVoteManipulation/{\num{0.2}}
\def\PActorTargetMainTechnicalVulnerabilityGeneralUndetermined/{\num{0.7}}
\def\PActorTargetMainTechnicalVulnerabilityGeneralUndeterminedSpecificUndetermined/{\num{0.7}}
\def\PActorTargetMainUndetermined/{\num{2.3}}
\def\PActorTargetMainUndeterminedGeneralUndetermined/{\num{2.3}}
\def\PActorTargetMainUndeterminedGeneralUndeterminedSpecificAccessingPrivateKeysData/{\num{2.3}}
\def\PActorTargetMainHumanRisk/{\num{3.1}}
\def\PActorTargetMainHumanRiskGeneralExternalFactor/{\num{1.5}}
\def\PActorTargetMainHumanRiskGeneralExternalFactorSpecificDeceivingPersonnel/{\num{0.8}}
\def\PActorTargetMainHumanRiskGeneralExternalFactorSpecificExploitingOperationalMistake/{\num{0.8}}
\def\PActorTargetMainHumanRiskGeneralInternalTheft/{\num{1.5}}
\def\PActorTargetMainHumanRiskGeneralInternalTheftSpecificBackdoor/{\num{0.2}}
\def\PActorTargetMainHumanRiskGeneralInternalTheftSpecificUnauthorizedUseofPrivateKey/{\num{0.8}}
\def\PActorTargetMainHumanRiskGeneralInternalTheftSpecificContractVulnerabilityExploit/{\num{0.2}}
\def\PActorTargetMainHumanRiskGeneralInternalTheftSpecificMaliciousCodeInjection/{\num{0.3}}
\def\PActorTargetMainHumanRiskGeneralInternalTheftSpecificUndetermined/{\num{0.1}}
\def\PActorPerpetrator/{\num{41}}
\def\PActorPerpetratorMainMaliciousUseofContract/{\num{35.9}}
\def\PActorPerpetratorMainMaliciousUseofContractGeneralRugPullScam/{\num{35.9}}
\def\PActorPerpetratorMainMaliciousUseofContractGeneralRugPullScamSpecificSellingRestrictions/{\num{4.8}}
\def\PActorPerpetratorMainMaliciousUseofContractGeneralRugPullScamSpecificLiquidityRemoval/{\num{16.3}}
\def\PActorPerpetratorMainMaliciousUseofContractGeneralRugPullScamSpecificHiddenMintFunction/{\num{3.5}}
\def\PActorPerpetratorMainMaliciousUseofContractGeneralRugPullScamSpecificPumpDump/{\num{6.5}}
\def\PActorPerpetratorMainMaliciousUseofContractGeneralRugPullScamSpecificUndetermined/{\num{4.8}}
\def\PActorPerpetratorMainMarketManipulation/{\num{4.9}}
\def\PActorPerpetratorMainMarketManipulationGeneralMisappropriationofFunds/{\num{4.9}}
\def\PActorPerpetratorMainMarketManipulationGeneralMisappropriationofFundsSpecificScamPresaleIDOandICO/{\num{4.5}}
\def\PActorPerpetratorMainMarketManipulationGeneralMisappropriationofFundsSpecificPonziScheme/{\num{0.2}}
\def\PActorPerpetratorMainMarketManipulationGeneralMisappropriationofFundsSpecificEmbezzlement/{\num{0.1}}
\def\PActorPerpetratorMainMarketManipulationGeneralMisappropriationofFundsSpecificUndetermined/{\num{0.1}}
\def\PActorIntermediary/{\num{7.1}}
\def\PActorIntermediaryMainImitation/{\num{7.1}}
\def\PActorIntermediaryMainImitationGeneralInstantUserDeception/{\num{7.1}}
\def\PActorIntermediaryMainImitationGeneralInstantUserDeceptionSpecificEvilTwinSite/{\num{0.4}}
\def\PActorIntermediaryMainImitationGeneralInstantUserDeceptionSpecificSocialMediaCompromission/{\num{3.1}}
\def\PActorIntermediaryMainImitationGeneralInstantUserDeceptionSpecificFakeAdsPopUps/{\num{0.9}}
\def\PActorIntermediaryMainImitationGeneralInstantUserDeceptionSpecificDNSAttack/{\num{1.3}}
\def\PActorIntermediaryMainImitationGeneralInstantUserDeceptionSpecificScamAirdrops/{\num{0.4}}
\def\PActorIntermediaryMainImitationGeneralInstantUserDeceptionSpecificUndetermined/{\num{0.4}}
\def\PActorIntermediaryMainImitationGeneralInstantUserDeceptionSpecificFakeService/{\num{0.2}}
\def\PActorIntermediaryMainImitationGeneralInstantUserDeceptionSpecificPhishingEmails/{\num{0.2}}
\def\PActorIntermediaryMainImitationGeneralInstantUserDeceptionSpecificFrontEndAttack/{\num{0.2}}
\def\NTarget/{\num{516}}
\def\NPerpetrator/{\num{425}}
\def\NIntermediary/{\num{50}}
\def\PTargetStackP/{\num{54}}
\def\NTargetStackP/{\num{282}}
\def\PTargetStackDLT/{\num{6.1}}
\def\NTargetStackDLT/{\num{32}}
\def\PTargetStackCA/{\num{17.8}}
\def\NTargetStackCA/{\num{93}}
\def\PTargetStackCP/{\num{14.9}}
\def\NTargetStackCP/{\num{78}}
\def\PTargetStackINT/{\num{7.1}}
\def\NTargetStackINT/{\num{37}}
\def\NActorTargetFinD/{\num{433}}
\def\AActorTargetFinD/{\$7.7B}
\def\NTargetTechnicalVulnerability/{\num{383}}
\def\ATargetTechnicalVulnerability/{\$6.4B}
\def\NTargetUndetermined/{\num{22}}
\def\ATargetUndetermined/{\$321M}
\def\NTargetHumanRisk/{\num{28}}
\def\ATargetHumanRisk/{\$980M}
\def\NActorPerpetratorFinD/{\num{405}}
\def\AActorPerpetratorFinD/{\$1.6B}
\def\NPerpetratorMaliciousUseofContract/{\num{356}}
\def\APerpetratorMaliciousUseofContract/{\$644M}
\def\NPerpetratorMarketManipulation/{\num{49}}
\def\APerpetratorMarketManipulation/{\$905.6M}
\def\NStackP/{\num{298}}
\def\AStackP/{\$2.6B}
\def\NStackCA/{\num{375}}
\def\AStackCA/{\$1.8B}
\def\NStackCP/{\num{107}}
\def\AStackCP/{\$925M}
\def\NStackDLT/{\num{24}}
\def\AStackDLT/{\$1.5B}
\def\NStackINT/{\num{34}}
\def\AStackINT/{\$2.5B}

%% file: sections/abstract.tex
% !TeX root = ../main.tex

\section*{Abstract}

Decentralized finance (DeFi) has been the target of numerous profit-driven crimes, but the prevalence and cumulative impact of these crimes have not yet been assessed. This study provides a comprehensive assessment of profit-driven crimes targeting the DeFi sector. We collected data on 1141 crime events from 2017 to 2022. Of these, 1,036 were related to DeFi (the main focus of this study) and 105 to centralized finance (CeFi). {The findings show that the entire cryptoasset industry has suffered a minimum loss of US\$30B, with two thirds related to CeFi and one third to DeFi.} Focusing on DeFi, a taxonomy was developed to clarify the similarities and differences among these crimes. All events were mapped onto the DeFi stack to assess the impacted technical layers, and the financial damages were quantified to gauge their scale. The results highlight that during an attack, a DeFi actor (an entity developing a DeFi technology) can serve as a direct target (due to technical vulnerabilities or exploitation of human risks), as a perpetrator (through malicious uses of contracts or market manipulations), or as an intermediary (by being imitated through, for example, phishing scams). The findings also show that DeFi actors are the first victims of crimes targeting the DeFi industry: \PActorTarget/\%  of events targeted them, primarily due to technical vulnerabilities at the protocol layer, and these events accounted for 83\% of all financial damages. Alternatively, in \PActorPerpetrator/\% of events, DeFi actors were themselves malicious perpetrators, predominantly misusing contracts at the cryptoasset layer (e.g., rug pull scams). However, these events accounted for only 17\% of all financial damages. The study offers a preliminary assessment of the size and scope of crime events within the DeFi sector and highlights the vulnerable position of DeFi actors in the ecosystem.

%% file: sections/introduction.tex
% !TeX root = ../main.tex

\section*{Introduction}

The rapid growth in value of cryptoassets~\footnote{This study uses the term ``cryptoassets'' to describe all digital representations of value that utilize a form of distributed ledger technology and can be transferred, stored, or traded electronically.}, coupled with recent developments in blockchain technologies, has provided momentum for the decentralized finance (DeFi) industry to emerge. The DeFi industry aims to provide transparent, open-source, and permissionless online financial services~\cite{zetzsche_decentralized_2020}. The term ``DeFi'' was coined in 2018 by entrepreneurs developing the Ethereum blockchain and ``refers to financial services that build upon the decentralized foundations of blockchain technology''\cite{a_deshmukh_blockchain_2022}. In essence, DeFi applications offer financial services such as swapping assets via decentralized exchanges (DEXs), lending and borrowing assets through lending protocols, and speculating on future prices using derivative products. Moreover, DeFi enables a novel form of software-driven financial engineering, allowing service providers to merge functions from various DeFi protocols to introduce novel, intricate, and deeply nested financial services. For a comprehensive overview of the technical building blocks and financial functions of DeFi, readers are directed to~\cite{auer_technology_2023}.

Decentralized finance is part of the cryptoasset industry, which also includes centralized finance (CeFi). In contrast to DeFi, CeFi allows the flexible trading of both fiat and cryptoassets through a centralized governance system~\cite{qin_cefi_2021}. Since the inception of the cryptoasset industry, government agencies have faced difficulties in developing and/or enforcing a regulatory framework surrounding the industry~\cite{zetzsche_decentralized_2020}. Moreover, the large amount of money that the industry handles, close to USD 2.3 trillion in 2021~\cite{noauthor_top_nodate}, coupled with its decentralized and pseudo-anonymous features, makes it an attractive setting for profit-driven crime~\cite{barone_cryptocurrency_2019, hendrickson_cash_2022, pieroni_smarter_2018}. Indeed, previous studies have shown that cryptoassets can serve as a means for money laundering~\cite{barone_cryptocurrency_2019, nolasco_braaten_convenience_2021, pieroni_smarter_2018} or as a means of payment for criminal activities such as extortion schemes~\cite{huang_geography_2018, paquet-clouston_ransomware_2019, paquet-clouston_spams_2019}. Within this industry, the DeFi sector specifically has also experienced its share of criminal events. For example, there have been several accounts of developers abandoning their DeFi projects and fleeing with their investors’ money, also known as rug pull scams~\cite{mackenzie_criminology_2022}. A vast array of techniques has also been developed to exploit blockchain technologies or smart contracts for theft purposes~\cite{chen_survey_2020}.

To date, several studies have explored specific crime types occurring in the DeFi industry~\cite{cao_flashot_2021,alshater_initial_2023, zhou_dependability_2021} or have developed tools to detect such attacks and/or establish countermeasures~\cite{wang_impact_2022, wang_blockeye_2021,nghiem_detecting_2021}. However, there is still a need to draw a broader picture of the crime landscape beyond technical specifics and crime types. Such a picture can inform future users, investors, and policymakers about where to focus their efforts and resources to better secure this emerging sector.

This study offers a comprehensive assessment of profit-driven crimes targeting the DeFi sector. To achieve this, we collected \NallDb/ crime events from 2017 to 2022, of which \NNonCeFi/ related to the DeFi industry and \NCeFi/ to the centralized finance (CeFi) industry. The first objective \textbf{(Obj. 1)} of the study is to develop an evidence-based taxonomy to identify the tactics and strategies used to steal money, as well as how DeFi actors are involved in crime events. In this study, DeFi actors represent any individual, group, or organization supporting the operations of DeFi services, including blockchain, fungible token (FT), non-fungible token (NFT), exchange, lending, derivative, Dapp, yield, staking, bridge, and oracle services. The second objective \textbf{(Obj. 2)} employs the framework developed by~\cite{auer_technology_2023} to assess which technical layer of the tech stack is most impacted by such crime events. The third objective \textbf{(Obj. 3)} quantifies the financial damages associated with these crimes. In short, the first objective provides a novel aggregate view of the crime landscape beyond technical specifics or broad categorizations like ``scam'' or ``fraud'', while the second objective allows for a clearer understanding of where malicious activities occur within the DeFi tech stack, and the third objective determines which types of crimes are most detrimental in the industry. To meet these objectives, a set of mixed methods, from content analyses to non-parametric tests, is employed. Our findings can be summarized as follows:

\begin{itemize}

\item {The cryptoasset industry suffered a minimum loss of \$30 billion from 2017 to 2022, with two thirds relating to CeFi and one third to DeFi.}

\item During an attack, DeFi actors can serve as a direct target due to technical vulnerabilities or human risk exploitation, become perpetrators through misuse of contracts or market manipulations, or act as intermediaries by being imitated through, for example, phishing campaigns. 

\item DeFi actors are the first victims of crimes targeting the DeFi industry: \PActorTarget/\% of crime events targeted them, mainly exploiting their technical vulnerabilities. These events accounted for nearly 83\% of all recorded financial damages.

\item A total of \PActorPerpetrator/\% of crime events were initiated by DeFi malicious perpetrators through misuse of contracts (e.g., rug pull scams). However, these events accounted for only 17\% of all financial damages.

\item The most impacted layers of the tech stack, in terms of number of crime events, were the DeFi protocol and Cryptoasset layers. The most financial damages were, however, experienced on the DeFi protocol and Interface layers.

\end{itemize}

These key insights highlight that the CeFi sector experiences larger financial losses. Also, many of attacks target legitimate DeFi actors who genuinely contribute to the ecosystem. Given that the sector does not yet have a reliable safety net~\cite{auer_technology_2023}, such as a supervision authority that systematically audits DeFi protocols or state-backed deposit insurance, it is crucial to protect actors from criminal conduct or fraud. This can be done by fostering collaboration between regulators and actors to establish clear guidelines for DeFi operations. Events where DeFi actors behave maliciously are also prevalent, but result in fewer financial damages. Although these events are smaller in scale, increasing user awareness that such malicious projects exists, while providing tools that can help them assess the potential risks of investing in a project should be advocated by all stakeholders in the ecosystem. The dataset developed for this study is available~\footnote{https://zenodo.org/records/14047933} to support research reproducibility and further analysis by other researchers.

%% file: sections/literature.tex
% !TeX root = ../main.tex

\section*{Crime in the DeFi Industry: A Review of the Literature}

Several types of crimes have been accounted for in the ecosystem, from Ponzi schemes~\cite{bartoletti_dissecting_2020,liang_data-driven_2021, gridley_significant_2023} to oracle attacks~\cite{wang_blockeye_2021,caldarelli_blockchain_2021}, rug pulls~\cite{puggioni_crypto_2022,xia_characterizing_2020} or scams~\cite{gridley_significant_2023,phillips_tracing_2020}. The review below provides an aggregated overview of studies focusing on crime in the decentralized finance industry. 

To this date, three studies have developed an aggregate analysis of crime events taking place in the DeFi ecosystem~\cite{reddy_cryptocurrency_2018,kris_oosthoek_flash_nodate,werner_sok_2022}. The author of one study~\cite{reddy_cryptocurrency_2018} argued that DeFi technologies could either be a facilitating tool for cybercrime (e.g., scam or fraud) or a target of attacks. When a target for attack, another study (which analyzed 20 reported vulnerability exploits) stated that DeFi actors could be victims of both protocol and market attacks~\cite{kris_oosthoek_flash_nodate}. Protocol attacks take advantage of a protocol implementation’s shortcomings and impacts said protocol only, while market attacks rather benefit from exploiting a protocol’s flawed business logic that can impact multiple protocols. For example, a protocol’s governance system choice could lead to a protocol attack, while manipulating an oracle could penalize multiple protocols, making it a market attack. This categorization also resembles the one developed in~\cite{werner_sok_2022} that separates technical security from economic security, as technical exploits on DeFi actors leave no room for the underlying blockchain system, markets, or other agents to react to such exploits, while economic exploits allow a window of reaction. In fact, the outcome to a smart contract exploit attempt is dichotomic, as either a profit will instantly be made or the operation will simply not be executed, while attempting to manipulate an oracle's or an asset’s price will typically require more steps, more capital, and won't have a guaranteed outcome~\cite{werner_sok_2022}. 

Hence, the literature differentiates between technical/security attacks aiming directly at the technology and market/economic attacks which relate to manipulating economic incentives. Beyond this categorization, scholars have conducted studies to systematize knowledge on the types of attacks taking place in the ecosystem and their related techniques~\cite{atzei_survey_2017,wen_attacks_2021,cao_survey_2022,chen_survey_2020}. 
These studies identify vulnerabilities on different layers of the blockchain, discuss the possible attacks to exploit them, state the consequences, and discuss defense mechanisms~\cite{atzei_survey_2017,wen_attacks_2021,cao_survey_2022,chen_survey_2020}. While they help understand how attacks work, they do not inform on the current state of the ecosystem, as they mainly rely on previous literature rather than real life cases. One exception includes a study by\cite{li_survey_2022}, which relied on real life events but did not elaborate on their prevalence or the extent of their financial damages. In addition, another study categorized the types of attack by blockchain layers by relying on a set of 181 reported DeFi crime events, existing literature, and audit reports~\cite{zhou_sok_2022}. While the framework presented remains very technical, the authors were able to provide insights on the ecosystem, as they found that the majority of DeFi incidents took place in late 2020 and peaked in August 2021~\cite{zhou_sok_2022}. They also stated that the most targeted blockchain layers were the smart contract and application ones~\cite{zhou_sok_2022}. This aligns with other studies pointing out that recurring exploitations include smart contract exploits and design flaws~\cite{ghaleb_towards_2022,qian_smart_2022,wen_attacks_2021}, such as flash loans being used as a vehicle to significantly amplify attack profit~\cite{kris_oosthoek_flash_nodate,qin_attacking_2021}. In their conclusion,~\cite{zhou_sok_2022} also estimated that the DeFi industry, including users, liquidity providers, speculators, and protocol operators have suffered a loss of at least 3.24 billion USD (p.1). 

Moreover, some studies account for the role of humans in technical/security attacks\cite{chen_survey_2020, li_security_2022, anita_blockchain_2019}. For example, one study proposed a taxonomy of vulnerability root causes which included flawed smart contract programming, Ethereum’s design/implementation and human factors~\cite{chen_survey_2020}. The authors introduced the idea that human factors, like improper configuration, can cause vulnerabilities, as it can lead to erroneous permissions for an Ethereum client. Another study highlighted that phishing attacks on team members and improper key management, such as inefficient storage methods, can lead to deployers’ private key compromission and their subsequent usage by unauthorized parties~\cite{li_security_2022}. In addition, \cite{anita_blockchain_2019} discussed insider attacks, which can occur when someone with administrative privileges accesses the computer system and performs unauthorized operations~\cite{anita_blockchain_2019}. 

DeFi users, moreover, represent another human vulnerability and several studies have investigated how DeFi users can be deceived through various fraudulent techniques, from social engineering~\cite{matakovic_crypto-assets_2022,andryukhin_phishing_2019} to phishing~\cite{s_shukla_addressing_2022} and more~\cite{ramos_great_2021,phillips_tracing_2020}. Two studies have also examined how traditional financial market frauds, such as Ponzi schemes and pump-and-dumps have modernized with cryptoassets: they can now be pre-programmed in smart contracts~\cite{kleinberg_cryptocurrencies_2022, bartoletti_dissecting_2020}. Additional studies have also investigated the various forms of scams taking place in the DeFi ecosystem~\cite{kamps_moon_2018,phillips_tracing_2020,xia_characterizing_2020}. For example, the prevalence of phishing scams was investigated by ~\cite{xia_characterizing_2020} which identified 300 fake exchanges apps and 1,595 scam domain. The uncovered fake applications affected a total of 38 legitimate exchanges, which included almost all major cryptoasset exchanges. Another study explained that some phishing exchanges were solely created to get users to voluntarily deposit assets, while others were created to replicate a wallet extension and steal users’ private keys~\cite{kamps_moon_2018}. Other studies also investigated imitation scams or giveaway scams~\cite{phillips_tracing_2020, kamps_moon_2018}, where an attacker typically imitates a celebrity on social media, claiming to giveaway cryptoassets. A study found that the average transaction value sent by users to different forms of giveaway scams ranged from USD 300  to USD 1,312~\cite{phillips_tracing_2020}. While all these scams are still ongoing, recent work shows that rug pull scams, where creators flee with investors' assets, are the latest rising fraudulent activity trend in the ecosystem, and one of the most financially damaging~\cite{puggioni_crypto_2022, chainalysis_2022_2022, xia_trade_2021}. Specifically, one study ~\cite{chainalysis_2022_2022} stated that users lost about USD 2.8B when rug pulls were at their peak in 2021. Also note that trading pairs and liquidity pools are central to these rug pull scams, as removing liquidity or dumping assets is often what initiates the exit scam~\cite{xia_trade_2021}. On the other hand,~\cite{xia_trade_2021} explained that the contract can be malicious in itself, as deployers hide minting functions or functions to restrict investors to sell their tokens. Some authors differentiate hard rug pulls, which refers to the use of a malicious smart contract to defraud investors, and soft rug pulls, which refer to asset manipulation by the developer, like pump-and-dump~\cite{puggioni_crypto_2022,xia_trade_2021}.

Finally, techniques have been developed to detect various types of crime events~\cite{la_morgia_pump_2020,mazorra_not_2022,xia_trade_2021}. For example, pump-and-dump scams were investigated by uncovering DeFi projects anomalies with machine learning~\cite{la_morgia_pump_2020,mansourifar_hybrid_2020}. Other studies have developed techniques to detect malicious contracts by scanning their code~\cite{mazorra_not_2022,xia_trade_2021}. For example, by categorizing more than 20,000 tokens on the Uniswap protocol, ~\cite{mazorra_not_2022} labeled 631 tokens as non-malicious, and 26,957 as malicious, suggesting that fraudulent tokens were more prevalent than legitimate ones in the DeFi ecosystem. 

\subsection*{This Study}

% previous work
Overall, to better understand crime trends in the DeFi industry, many studies either inform on the types of attack taking place in the ecosystem~\cite{atzei_survey_2017,chen_survey_2020}, explore a type of attack thoroughly~\cite{cao_flashot_2021,alshater_initial_2023, zhou_dependability_2021} and/or develop tools to detect such attacks and establish countermeasures~\cite{wang_impact_2022, wang_blockeye_2021,nghiem_detecting_2021}. Given this review, one can see that the DeFi industry represents a hotbed for crime. To better understand crime trends in the DeFi ecosystem, categorizations have been developed, from economic/market~\cite{werner_sok_2022, kris_oosthoek_flash_nodate} to technical (e.g, contract vulnerability exploitation)~\cite{zhou_sok_2022} to social attacks (e.g., scams)~\cite{matakovic_crypto-assets_2022,s_shukla_addressing_2022}. Several specific types of crimes have also been investigated~\cite{bartoletti_dissecting_2020, puggioni_crypto_2022}.

% gap
However, as of today, there is a need to better understand crime similarities and differences targeting the DeFi industry, beyond broad categorizations and/or an exclusive focus on their technical details. Moreover, although the method used and the types of crimes are well known, what common strategies are used to steal money has yet to be uncovered and analyzed. Differentiating these aspects in crime events and assessing their prevalence and level of financial damages can inform policy makers on where to mitigate financial risks. Indeed, if the most prevalent and damaging events are those where the DeFi actor is, for example, complicit to the crime, then policy makers should develop better auditing processes to verify the legitimacy of DeFi actors as well as campaigns to better inform customers. On the other hand, if DeFi actors are most often targeted by external actors, then policy makers should encourage cybersecurity investments to better protect them.

% our contribution
Hence, this study provides a first comprehensive assessment of profit-driven crimes targeting the DeFi sector using \NallDb/ reported crime events from 2017 to 2022. For clarity, we use Naylor's general theory of profit-driven crime~\cite{naylor_towards_2003, naylor_predators_2003} to define what crime events are of interest to this study. Specifically, we focus on crime events that involve the illegal redistribution of existing wealth from a victim to an offender, also known as predatory crimes~\cite{naylor_towards_2003}. We do so by relying on reported events in the media, as these crimes are typically reported by victims~\cite{naylor_towards_2003, naylor_predators_2003}. To better understand the wide array of predatory profit-driven crimes taking place in this sector, we develop an evidence-based taxonomy and identify the tactics and strategies used to steal money, as well as how DeFi actors are involved in crime events (\textbf{Obj.1}). We also map which layer is most impacted by such crimes, drawing from DeFi Stack Reference Model (DSR) proposed by~\cite{auer_technology_2023} (\textbf{Obj. 2}) and we quantify the financial damages associated with these crimes (\textbf{Obj. 3}).

%% file: sections/data_methods.tex
% !TeX root = ../main.tex

\section*{Data and Methods}

To collect crime events targeting DeFi actors, we relied on three aggregators: 1)~De.Fi~\cite{noauthor_fi_nodate}, 2) SlowMist~\cite{noauthor_slowmist_nodate-1}, and 3) CryptoSec (now ChainSec)~\cite{noauthor_cryptosec_2021}. De.Fi is a platform that ``aims to create a safe place for everyone to securely access DeFi services and protect users from the ecosystem’s potential risks''~\cite{defi_announcing_2023}. According to their website, they launched the REKT database in August 2021, which gathers information on reported and mediatized DeFi scams, hacks, and exploits, including the technical issues (if any) and lost funds. The organization behind De.Fi provided an API key so the REKT database data could be automatically fetched. SlowMist, on the other hand, is advertised as a ``blockchain security firm that provides various services such as security audits, threat information, bug bounties and defense-deployment''~\cite{noauthor_slowmist_nodate}. The firm publishes a list of crime events targeting many high-risk blockchain security flaws on a section labeled ``hacked'' on their websites. Finally, CryptoSec is an online resource focused on providing non-technical information about safeguarding assets from hacks and theft. ~\cite{noauthor_cryptosec_2021}. Their website contains a section listing crime events targeting DeFi actors. Note that as of February 2023, this aggregator is now called ChainSec~\footnote{https://chainsec.io/defi-hacks/}. For these two aggregators, we manually collected the information.

\subsection*{Dataset Creation}

Initially, all crime events listed by the aggregators from 2013 to 2022 were collected (De.Fi $N=2,699$, SlowMist $N=907$ and CryptoSec (now ChainSec) $N=106$), leading to 3,712 events. As event coverage greatly varies between aggregators, we used them as complementary rather then comparatively. We obtained the  majority of our information from De.Fi REKT database. This aggregator was prioritized because: 1) the authors behind the database built the database with a clear objective to identify and list all crime events,  2) they converted the stolen amount in USD at the time of the recorded event, and 3) they based their conclusion on a proactive blockchain investigation approach, providing links, most of the time, towards relevant transactions and addresses as proof of work. SlowMist and CryptoSec (now ChainSec), on the other hand, are mainly news aggregators and cover the most mediatized events.

During the Fall of 2022 and early 2023, one author was tasked with creating the crime event dataset. The events were coded according to the content of the event summary provided by the aggregator, which gathered basic information, such as how the event unfolded and who was targeted. As none of the aggregators publicly disclosed a thorough method on how they collected, reported and summarized events, measures were taken to alleviate the risk of relying on false information. First, events' occurrence and summaries' legitimacy were verified by consulting all sources linked by aggregators. Such sources included post-mortem analysis from involved DeFi actors and security firms, as well as news articles and relevant social media posts. Second, for less detailed summaries with no external links provided by the aggregators, we conducted searches on the Google search engine to find additional sources. Consulting external links from aggregators, as well as conducting manual searches, enabled us to verify the accuracy of the information and ensure a comprehensive understanding of the event. A first review of the dataset was conducted to categorize events and remove all crime events for which aggregators did not link sources, and no related sources manually could be found  ($N=73$).   

The dataset was reviewed by all four authors in an independent session. In this session, open discussions led to removing certain types of crime events. For instance, all crimes that were not profit-driven ($N=38$) were removed. We also removed all events that did not directly involve a DeFi actor ($N=30$), such as regular fiat currency thefts where assets were later laundered through DeFi services. In addition, all events discussing vulnerability discoveries through security teams or bug bounty programs were excluded ($N=19$), along with all attempted attacks ($N=6$). Events from which we did not have enough information to understand the actor's involvement in the crime event were also dismissed ($N=15$). 

Moreover, some DeFi scams on De.Fi REKT were listed as a warning rather than as an event report ($N=1,859$). Precisely, De.Fi REKT listed these events as scam projects as they noticed their smart contract contained malicious terms or functions that could represent a risk for investors. Since no financial losses nor crimes had yet taken place, we removed them from the dataset. Finally, we merged all duplicates between and within aggregators ($N=480$), and removed  all crime events taking place prior to 2017 ($N=51$), as fewer events were listed for those earlier years, and limited information was available for such events. This resulted in a sample of \NallDb/ reported profit-driven crime events.

\subsection*{Information Extraction from Crime Events}

From these \NallDb/ crime events, we extracted the name of the DeFi actor involved in the crime event along with the date of the event and a description on the course of the event. Such descriptions, together with our review of additional sources, were used to create the taxonomy presented below. Considering the date of the event, 
Figure~\ref{fig:eventyear} shows the number of reported crime events in our dataset per year. We can observe an increase over the years, with more than half of the events occurring in 2021 and 2022. This increase can be explained by DeFi's rapid growth during that period. Note that in the past years, efforts have also been put in place to detect and report cryptoassets-related crimes, which might explain why more crime events have been uncovered recently.

{
	\begin{figure}
		\centering
		\includegraphics[width=0.9\columnwidth]{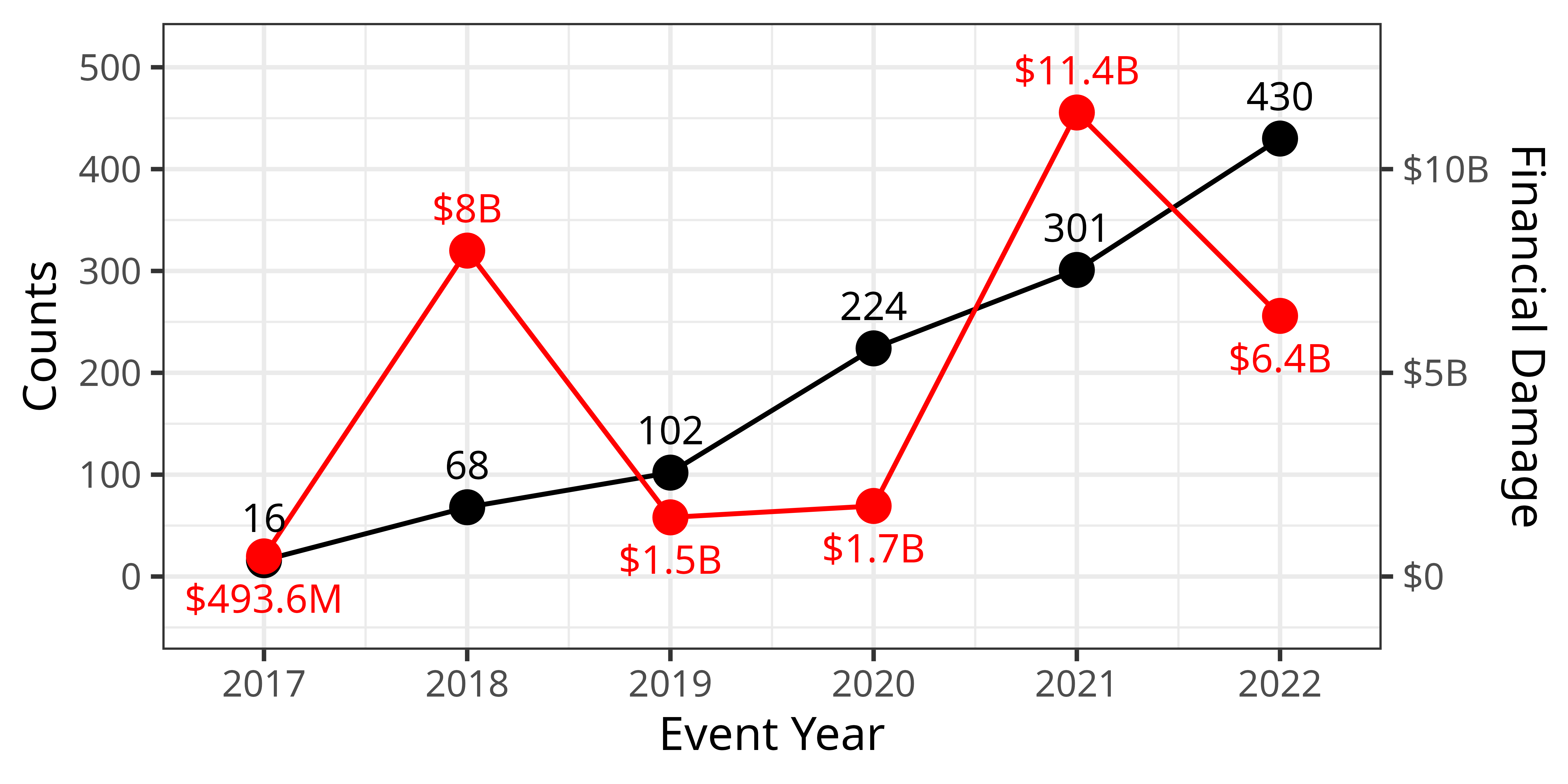}
		\caption{\textbf{Total number and Total Financial Damages of reported crime events per year}. This figure shows the total reported profit-driven crime events (black line) and damages (red line) over the observation period of our sample.}\label{fig:eventyear}
	\end{figure}
}	

To gather stolen amounts, we first used the information available on De.Fi's REKT database which reported USD amounts as the database curators conducted the conversion when reporting the crime~\cite{defi_announcing_2023}. For events retrieved solely on one of the two other aggregators, the reported amount was converted from the stolen currency to USD using CoinMarketCap’s historical data~\cite{noauthor_check_nodate}. For precision purposes, we used the mean of the four USD recorded values the website provides per date (earliest data in range, lowest, highest, and last). This value was then multiplied by the initial stolen currency's amount to obtain the stolen amount value in USD. Hence, all results below are in USD and the currency is not mentioned anymore when reporting amounts in the text. To double-check the information collected, we randomly selected 50 crime events (coming from the three aggregators) and verified the amounts to those reported by other sources. Specifically, we searched other news articles announcing the crime event and compared the amount in news articles (e.g., CoinDesk, Crypto News) with the one announced by the aggregator. The amount in our final dataset was consistent with other sources in 94\% of the time (47/50). Information on stolen amounts was available for \NallDbHasFinDam/ events.

Figure~\ref{fig:eventyear} also illustrates the total yearly financial damages reported in our dataset, showing a significant increase in 2021. In terms of descriptive statistics, the minimum stolen amount was \NMinFinDam/ and the maximum stolen amount was a non-negligible \NMaxFinDam/. The latter related to the compromission of the South African platform Africrypt~\cite{noauthor_south_2021}. On average, \NMeanFinDam/ was stolen per incident. The standard deviation was \NStdFinDam/. In at least 50\% of the crime events, more than \NMedianFinDam/ was stolen. The total amount stolen across these  \NallDbHasFinDam/ crimes was \NSumFinDam/. A summary of these statistics is presented in  Appendix~\ref{subsection:crimeevents}.

Note that throughout the first phase of the analysis, \NCeFi/ crime events related to centralized finance (CeFi) were identified in our dataset. CeFi has similarities to traditional forms of centralized finance as exchanges manage cryptoasset transactions for customers. DeFi, on the other hand, enables peer-to-peer transactions without the need for a centralized exchange. Yet, comparative estimates on CeFi and DeFi are still presented in the results section to provide an idea on the scale and scope of DeFi crime events, compared to CeFi, as they are part of the same cryptoasset industry. The final dataset on crime events targeting the \textbf{DeFi sector} thus includes \NDeFi/ DeFi crime events, with \NNonCeFiHasFinDam/ including a stolen amount.

\subsection*{Obj 1: Taxonomy Creation}

The evidence-based taxonomy was created by qualitatively analysing \NDeFi/ crime events that took place between 2017 and 2022. To create a taxonomy, we took a bottom-up approach and conducted an inductive content analysis~\cite{kyngas_inductive_2020}. This analysis involved deriving patterns or categories from the data. We first read all crime events and described them (open coding phase). From the description, we then created categories to assess the types of crimes and the methods used to steal money. These categories were then aggregated in larger categories (abstraction phase) with the aim to group crime events together based on how the money was stolen. This led us to develop three categories: \emph{specific tactics}, \emph{general tactics} and \emph{strategies}. 

Specific tactics referred to the technique used to realize the crime, often as reported in the literature, such as rollback attacks or Ponzi schemes. General tactics, on the other hand, focused on finding the common techniques or methods used by malicious actors to steal money, such as exploiting a contract or misappropriating funds. Such general tactics could further be aggregated in all-encompassing categories that displayed the strategy accounting for the main approach used to steal funds. This category denotes the high-level approach used in the illicit extraction of funds, such as exploiting a human risk or a technical vulnerability. Finally, based on the three previous categories, it seemed obvious that the implication of the DeFi actor in the crime event had to be differentiated. This differentiation provided a clearer understanding of their role in such events. The taxonomy is presented in the results section, along with crime prevalence in each category.

\subsection*{Obj 2: Mapping Crime Events on the Technical DeFi Stack}
\phantomsection
\label{subsec:method_obj2}
% Define a command for customized reference text
\newcommand{\customref}[2]{\hyperref[#1]{#2}}

To meet our second objective, we first identified the primary technical area of operation for DeFi actors within the DeFi sector. We then mapped this categorization to the DeFi Stack Reference Model (DSR) proposed by~\cite{auer_technology_2023} to determine where crime events had the most significant impact.

Using categorizations from crime aggregators and DeFiLlama~\cite{noauthor_defillama_nodate} --- the most extensive aggregator of DeFi actors to date --- as well as information from the actors' websites, we established 12 categories. These categories represent the primary technical areas of operation for DeFi actors: \emph{blockchain}, \emph{fungible tokens (FT}), \emph{non-fungible tokens (NFT)}, \emph{exchanges}, \emph{lending}, \emph{derivatives}, \emph{Dapps}, \emph{yield farming}, \emph{staking}, \emph{bridges}, \emph{oracles}, and \emph{others}. 
Figure~\ref{fig:areaevents} illustrates the total number of events and financial damages for each actor's category (which is based on their main area of operation) over the years. For the detailed numbers, see Appendix~\ref{subsection:crimeevents}.
Fungible tokens, Dapps and exchange services were most involved in crime events. However, events involving bridges or the underlying blockchain faced higher financial damages.
{
\begin{figure}
	\centering
	\includegraphics[width=0.9\columnwidth]{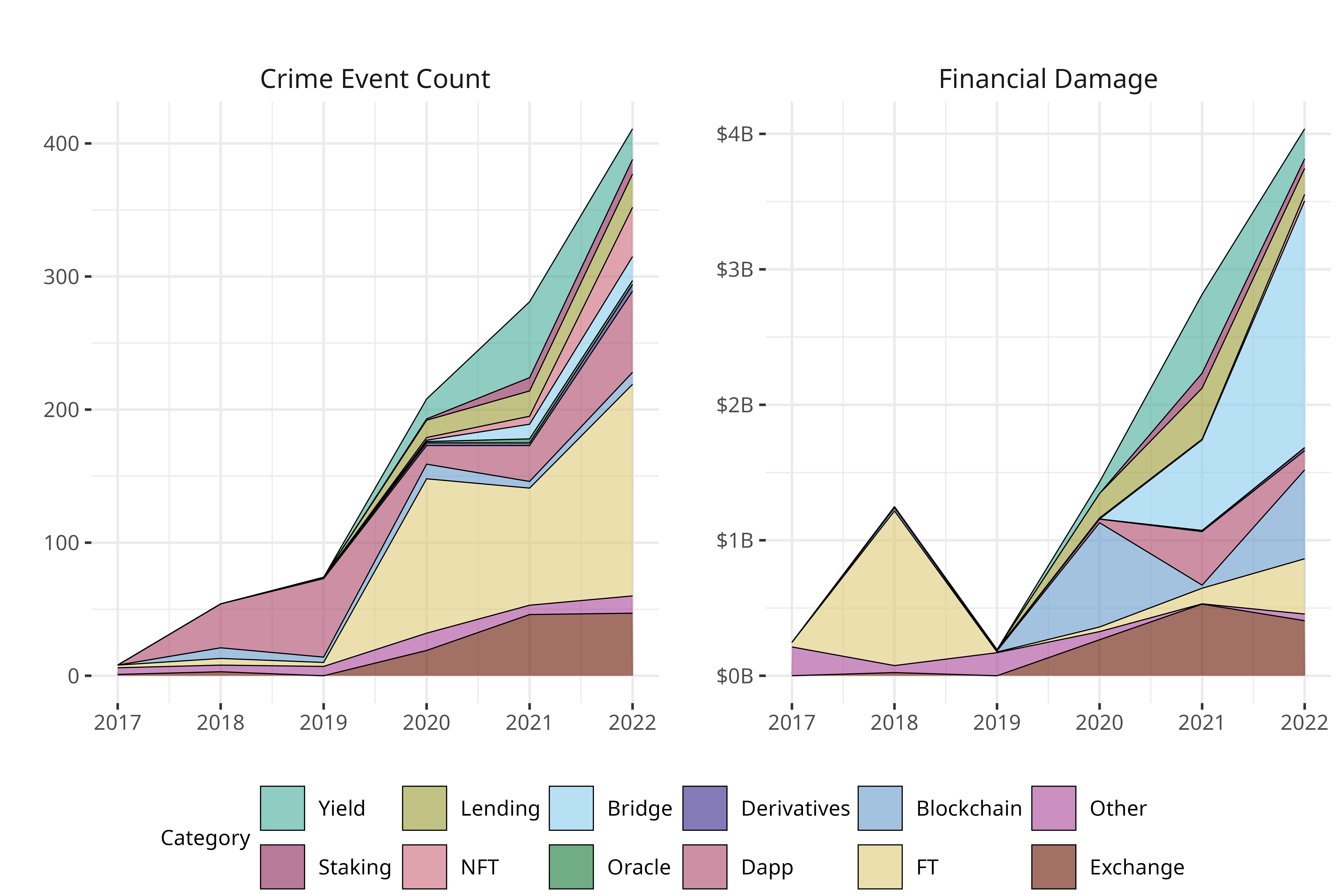}
	\caption{\textbf{Count and Financial Damages per Year per Actor Category.} This figure shows the total financial damages per year per Actor Category.}\label{fig:areaevents}
\end{figure}
}

% Explain enhanced DeFi stack model
Next, we aligned this categorization with the DSR model. A key attribute of this framework is the ``abstraction principle''. Each layer encompasses well-defined functions, utilizing functionality from the layer directly below and offering functionality to the one above. Since the DSR primarily represents DeFi protocols without accounting for the surrounding technical context, we augmented the model to include this aspect, situating it within a broader system environment.

\begin{figure}[t]
	\centering
	\includegraphics[width=8cm]{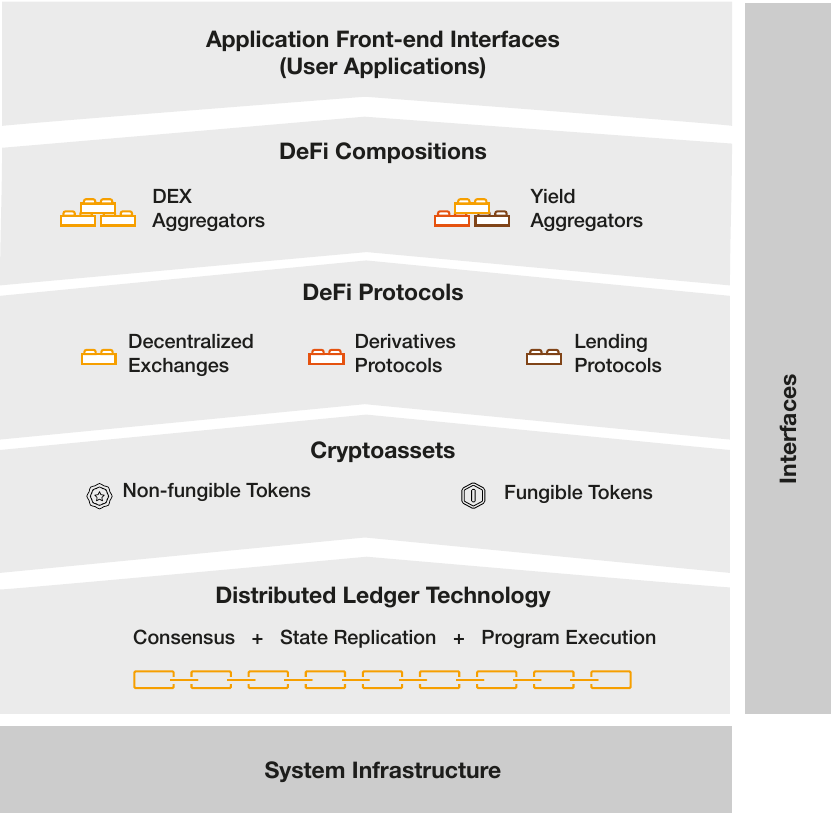}
	\caption{\textbf{Augmented DeFi Stack Reference Model (DSR).} Starting from the foundational \emph{System Infrastructure}, the model ascends through the \emph{Distributed Ledger Technology}, representing transaction settlements, to \emph{Cryptoassets}, symbolizing transferable value. \emph{DeFi Protocols} provide advanced services on cryptoassets and can be combined into \emph{DeFi Protocol Composition} providing new services. The top layer represents \emph{User Applications} and the orthogonal \emph{Interfaces} layer oracles and bridges. Layers from the original DSR are in light-gray, while augmented layers are drawn in dark-gray.}
	\label{fig:DeFi_stack_overview}
\end{figure}

Figure~\ref{fig:DeFi_stack_overview} presents the DSR model with its technical layers, embedded within a broader system environment. The \emph{System Infrastructure} layer forms the foundation of the DSR model. This layer can be targeted by crime events, potentially impacting all layers above it. Built upon the SYS layer, the DSR model introduces the \emph{Distributed Ledger Technology} layer, tasked with settling transactions and executing programs in a distributed system environment. The subsequent layer, \emph{Cryptoassets}, signifies assets that represent transferable value within a DLT system. \emph{DeFi Protocols} layer offers advanced financial services constructed on cryptoassets. Examples include Decentralized Exchanges that facilitate cryptoasset exchanges, Lending protocols enabling users to lend or borrow cryptoassets, and Derivative protocols that allow trading of synthetic positions mirroring the value of an underlying asset. A distinctive feature of DeFi is its ability to harness the financial functions of various DeFi protocols to introduce new financial services, termed as \emph{DeFi Protocol Composition}. \emph{User Applications} denote the pinnacle layer of the DSR model. Additionally, we have incorporated the \emph{Interfaces} layer—commonly known as Oracles and Bridges—as a distinct orthogonal layer.

Using this model, we identified the highest and most specific layer impacted in the DeFi stack model for each crime event in our sample, based on the actor's category. For instance, a crime event associated with a DeFi lending protocol would be allocated to the DeFi Protocols (DeFi P) layer. Considering the previously mentioned ``abstraction principle'', such an event would also encompass Cryptoassets (CA), as DeFi lending protocols enable the lending and borrowing of cryptoassets.

\subsection*{Obj 3: Measuring and Comparing Financial Damages}

To achieve the third and final objective, financial damages were measured and compared according to the taxonomy and tech stack presented before. Due to the highly skewed distributions, non-parametric tests were used, including Mann-Whitney U, Kruskall-Wallis and Dunns' post hoc tests. The significance level was set to 0.05 ($\alpha = 0.05$). When there were only two groups to compare, a Mann-Whitney U test was computed. The Mann-Whitney $U$ test compares the distributions of a continuous variable between two independent groups~\cite{shier_statistics_2004}. When significant, effect size were reported using cliff's delta ($d$)~\cite{meissel2024using}. When comparing multiple groups, Kruskal-Wallis H tests were computed~\cite{ostertagova_methodology_2014} and effect sizes were measured by calculating the epsilon squared ($\epsilon^2$)~\cite{tomczak2014need}. Given significant results of a Kruskall-Wallis H test, Dunn’s post hoc test with Bonferroni correction were finally computed~\cite{dunn_estimation_1959,dunn_multiple_1961-1}. Precisely, Dunn’s test identifies which groups are driving the significant difference found in the Kruskal-Wallis H test by comparing the difference between all pairs of groups. Since Dunn’s test computed each pairwise comparison separately, a Bonferroni correction was used to adjust the p-value and thus reducing the chances of doing a Type 1 error. Finally, effect sizes for all significant post-hoc tests were measured using the cliff's delta ($d$)~\cite{meissel2024using}. Also, when test results are presented, the median ($\Tilde{x}$) for each type of crime events is reported.

%% file: sections/results.tex
% !TeX root = ../main.tex

\section*{Results}

The results section below is organized into four subsections. First, we present CeFi and DeFi estimates, followed by the distribution of crime events according to the taxonomy we created. Next, we describe which layers in the technical stack are most affected by these crime events, followed by an assessment of their financial damages.

\begin{figure}[t]
	\centering
	\includegraphics[width=0.9\textwidth]{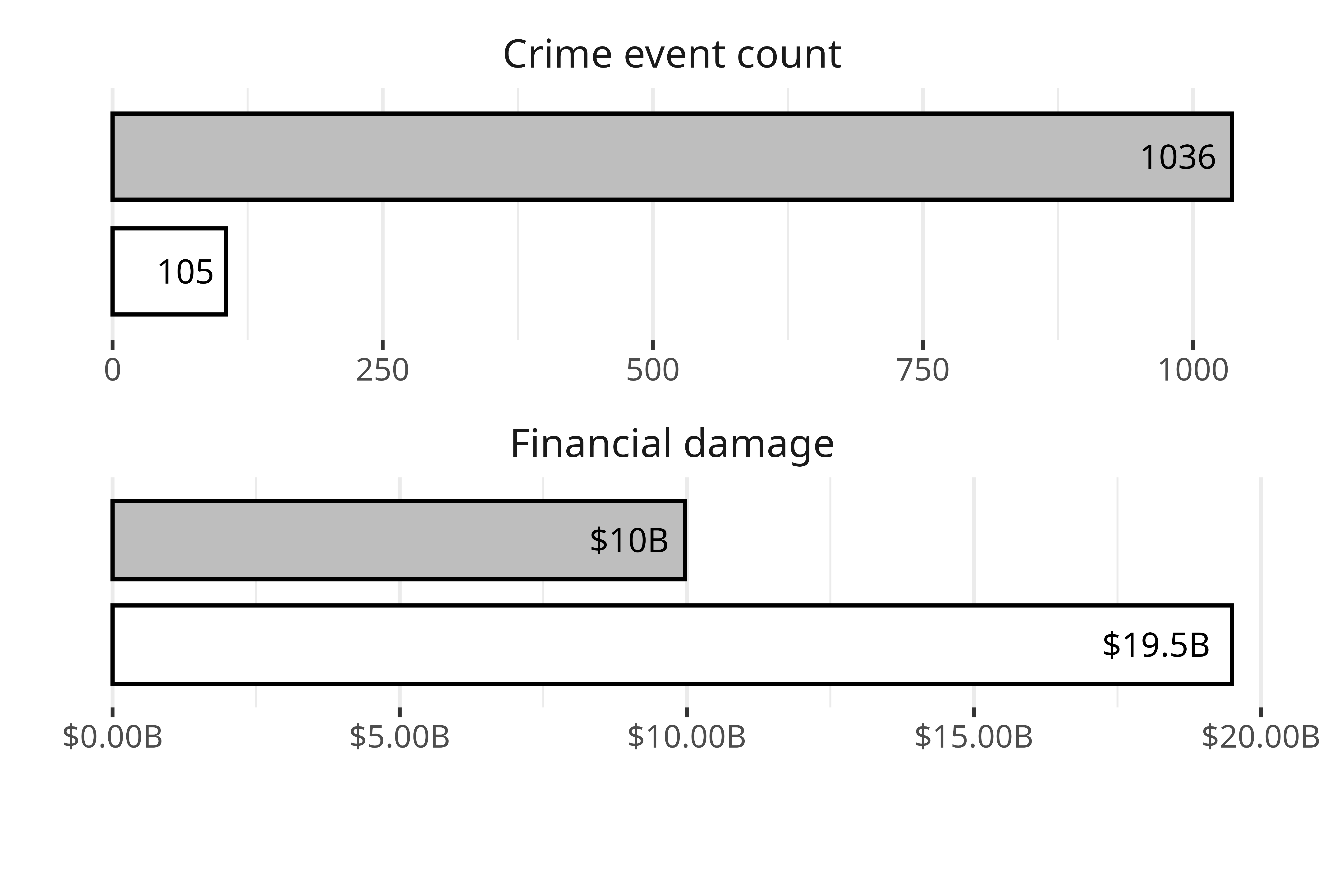}
	\caption{\textbf{Size and scope of DeFi crimes in comparison to CeFi.} This figure illustrates the contrast in terms of event prevalence, as well as financial damages observed between CeFi and DeFi.}
	\label{fig:CeFi_DeFi}
\end{figure}

\subsection*{CeFi vs. DeFi Estimates}

As explained in the section above, we identified that some crime events involved centralized finance (CeFi) actors ($N=\text{\NCeFi/}$), like centralized cryptoasset exchanges (CEX) that allow the flexible trading of both fiat and cryptoassets through a centralized governance system~\cite{qin_cefi_2021}. This finding is interesting as it shows the size and scope of DeFi crimes in comparison to CeFi, Figure~\ref{fig:CeFi_DeFi} illustrates the non-negligible role of CeFi in terms of financial damages related to cryptoassets. In fact, fewer events involved a CeFi actor, but they accounted for almost twice the financial damages as events involving a DeFi actor. One possible explanation is that high-scale fraud events, like Ponzi schemes and embezzlement, were mostly observed in the CeFi scene. Indeed, Ponzi schemes amounted to \NCeFiPonzi/  in stolen funds alone, and embezzlement schemes amounted to \NCeFiEmbezzlement/ in our dataset. However, it is important to note that the data collection process focused specifically on compiling a comprehensive dataset of DeFi crime events, rather than CeFi crime events. As a result, there are likely additional crime events targeting the CeFi industry that are not captured in this dataset. Consequently, these figures highlight the significantly higher financial losses in the CeFi sector compared to DeFi. 
%Further research could expand this study by incorporating CeFi crime events into the open-source dataset. 
Note that CeFi events are not included in further results and analyses, as the framework we developed is ultimately used to map events onto a DeFi-specific technical stack.

\subsection*{Introducing an Evidenced-based Taxonomy}

The first objective of this study is to develop an evidence-based taxonomy to identify the tactics and strategies used to steal money, as well as how DeFi actors are involved in crime events. The evidence-based taxonomy created thus illustrates how money can be --- and is --- stolen in the DeFi sector through a hierarchical structure that can be categorized along four main dimensions: (1) implication of DeFi actors, (2) strategies, (3) general tactics, and (4) specific tactics.

\begin{figure}
	\includegraphics[width=0.9\textwidth]{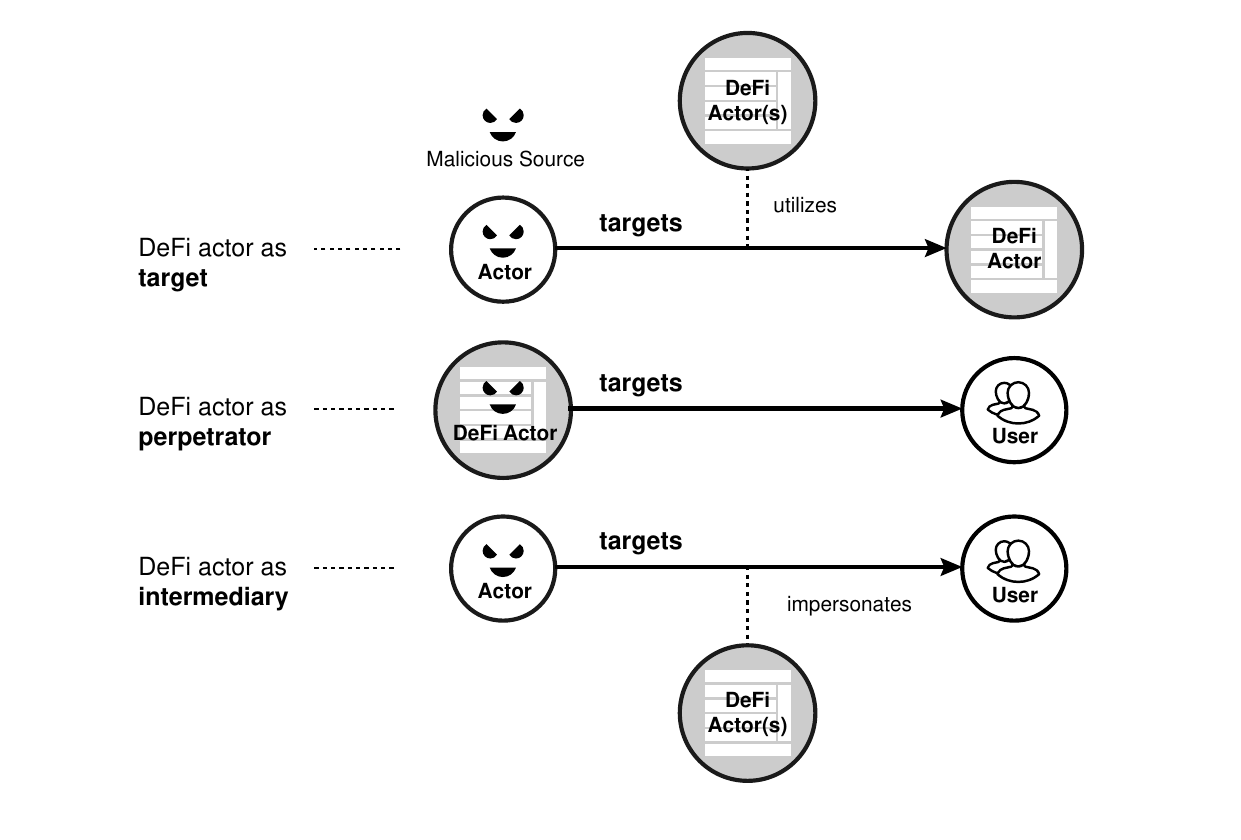}
	\caption{\textbf{Implications of DeFi actors in crime events.} This figure depicts the versatile roles DeFi actors can occupy in profit-driven crime through a simplified representation of a crime event unfolding, from the malicious source to the victimized party. Depending on the scenario, the implication of a DeFi actor ranges from being the victim, the malicious source itself, or an intermediary.}
	\label{fig:implications}
\end{figure}

To begin, Figure~\ref{fig:implications} highlights the possible implications of DeFi actors in crime events. DeFi actors can be direct targets of theft by a malicious actor, or they can themselves be 
perpetrators. In the latter case, DeFi actors and malicious actors merge into one actor targeting users. The third implication is an intermediary and, in this case, DeFi actors are used as intermediaries to reach users. Each implication is further explained below, considering the related strategies, as well as general and specific tactics uncovered. All statistics below represent proportions taking into account the \NDeFi/ DeFi-specific crime events in the dataset. For concision purposes, definitions on specific tactics are presented in Appendix~\ref{subsection:definitions}.

\subsubsection*{DeFi Actor as Target}

\begin{figure}
	\centering
	\includegraphics[width=0.9\textwidth]{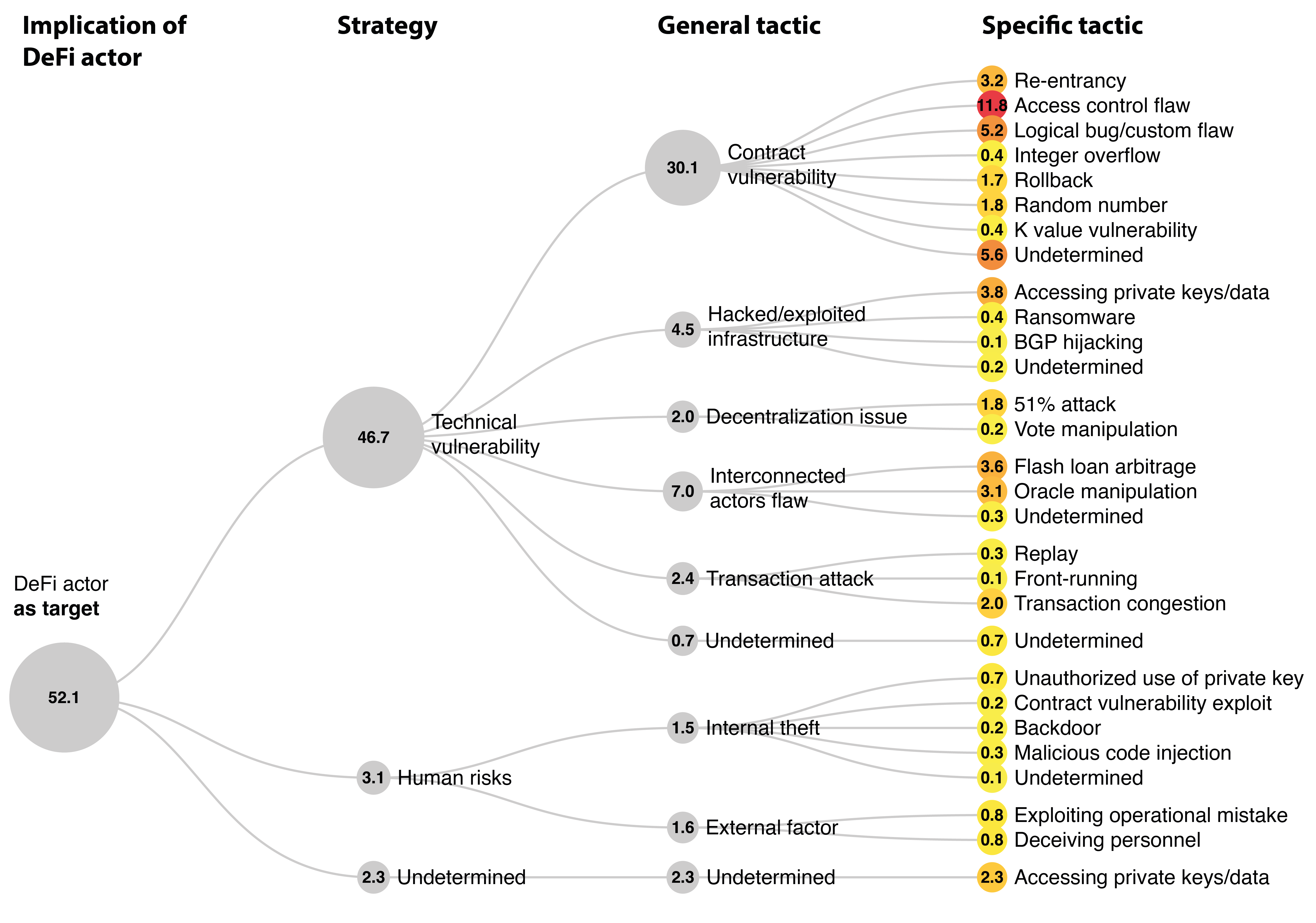}
	\caption{\textbf{DeFi Actor as Target: Distribution of Strategies and Tactics}. Percentages are calculated based on the entire dataset. Our taxonomy shows that for the first scenario (DeFi actor as target), the malicious actor uses strategies like the exploitation of technical vulnerabilities or human risks to victimize DeFi actors. Further desegregating such strategies into general and specific tactics then provides additional insight on how the event can be conducted.}
	\label{fig:tax_actor_target}
\end{figure}

Crime events in which DeFi actors were direct targets represented \PActorTarget/ \% of the dataset and could be regrouped under two overarching strategies: exploiting technical vulnerabilities (46.7\%) and exploiting human risks (3.1\%)~\footnote{Note that in some technical vulnerability exploits, a third party DeFi actor's service (flash loans, oracles) could also be leveraged by malicious actors to better target the desired DeFi actor.}. In some cases (2.3\%), we could not determine which strategy was used. The distribution of the strategies, general tactic and specific tactics are presented in Figure~\ref{fig:tax_actor_target}.

Specifically, contract vulnerability exploits included all events that aimed at exploiting smart contracts for malicious purposes. On the other hand, hacked/exploited infrastructure referred to crime events in which malicious actors successfully gained access to or exploited DeFi actors’ ``traditional” infrastructures, such as servers or corporate emails. Another tactic was exploiting interconnected actors' flaws which referred to situations where DeFi actors' interconnections created an opportunity for theft. Also, all crime events in which actors benefited from the way transactions were processed in the blockchain were classified as transaction attacks. Lastly, all crime events that exploited a vulnerability by taking advantage of consensus mechanism loopholes or governance systems were labeled as decentralization issues. The second strategy related to exploiting human weaknesses, limitations, errors, or trust. In the database, two general tactics were found: internal theft and external theft exploiting human factors. The first one referred to insiders committing profit-driven crimes by taking advantage of their strategic position in the DeFi actor’s organization to perform unauthorized operations for personal gains. The second referred to events created by an outsider that take advantage of humans, as opposed to technical features, to steal money.

Events for which we could not determine the strategy used were categorized as undetermined. This category should not be seen as different from the two previous ones, but rather as a category encompassing events for which it was unclear or undisclosed if the strategy was rooted in technical vulnerabilities or exploiting human factors.

\subsubsection*{DeFi Actor as Perpetrator}

\begin{figure}
	\centering
	\includegraphics[width=0.9\textwidth]{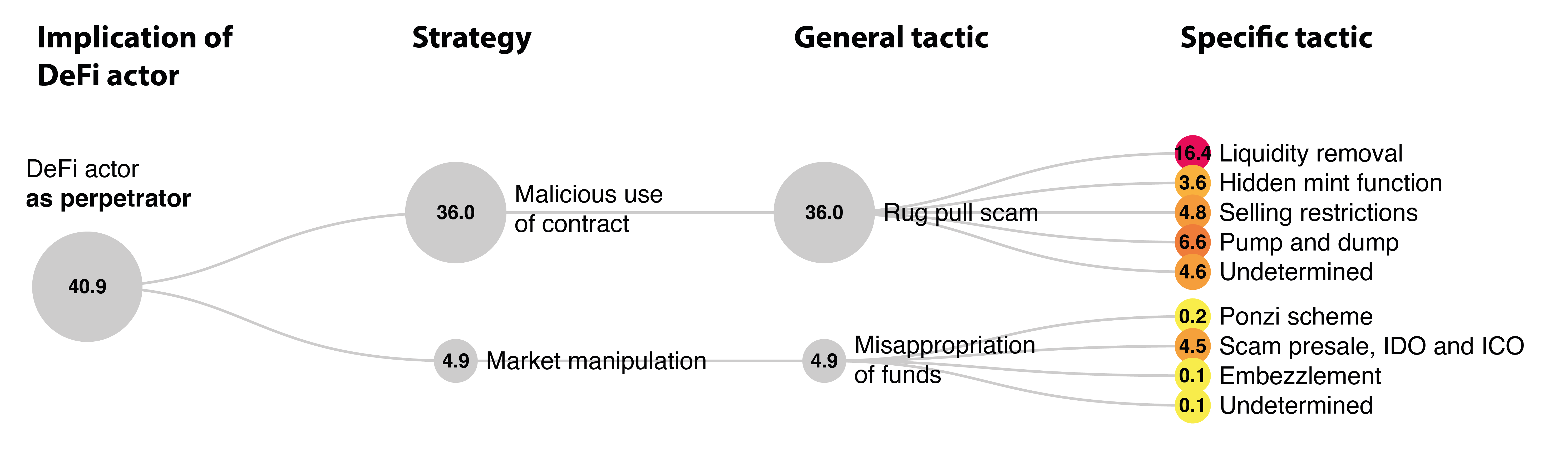}
	\caption{\textbf{DeFi Actor as Perpetrator: Distribution of Strategies and Tactics}. Percentages are calculated based on the entire dataset. Our taxonomy shows that for the second scenario (DeFi actor as perpetrator), the DeFi actor uses strategies like the malicious use of contracts and market manipulation to defraud users. Further desegregating such strategies into general and specific tactics then provides additional insight on how the event can be conducted.}
	\label{fig:tax_actor_malicious}
\end{figure}

Instead of being targets, DeFi actors were, in 40.9\% of the time, initiating the crime scheme. In this case, malicious actors and DeFi actors merged together, as the DeFi actor was not legitimate and aimed to directly steal users. This implication alluded to scams where a developer deliberately created and operated a DeFi project with the intent to steal its investors’ assets. We observed two strategies used by DeFi perpetrators to steal money: malicious use of smart contracts (36\%) and market manipulation (4.9\%), as shown in Figure~\ref{fig:tax_actor_malicious}. The first strategy, malicious use of contracts, referred to developers maliciously interacting with the contract they created through different operations to generate personal profit at the expense of other investors. Note that while existing literature differentiates between ``soft rug pulls" that involve removing or selling assets and ``hard rug pulls" that leverage malicious code within the contract itself, we categorized all these specific tactics under the term ``rug pull scam". For one, regardless of the specific tactic employed, the underlying strategy is the malicious utilization of a contract. Additionally, when examining the transaction history of contracts, it became evident that specific tactics can be used in complementary ways. For example, contract deployers could enable a hidden mint function prior to selling their share of assets.

The second strategy, market manipulation, referred to crime events in which DeFi perpetrators employed deceptive tactics to influence users to invest money in a project. These manipulative strategies led to illusory profitable trading or investment opportunities, which enabled perpetrators to profit at the expense of deceived participants. The only general tactic identified was misappropriation of funds. While rug pulls also rely on influencing users to invest in their token, misappropriation of funds does not require subsequent operations or malicious coding. Again, the distribution of the strategies, general tactic and specific tactics are presented in Figure~\ref{fig:tax_actor_malicious}.

\subsubsection*{DeFi Actor as Intermediary}

\begin{figure}[t]
	\centering
	\includegraphics[width=\textwidth]{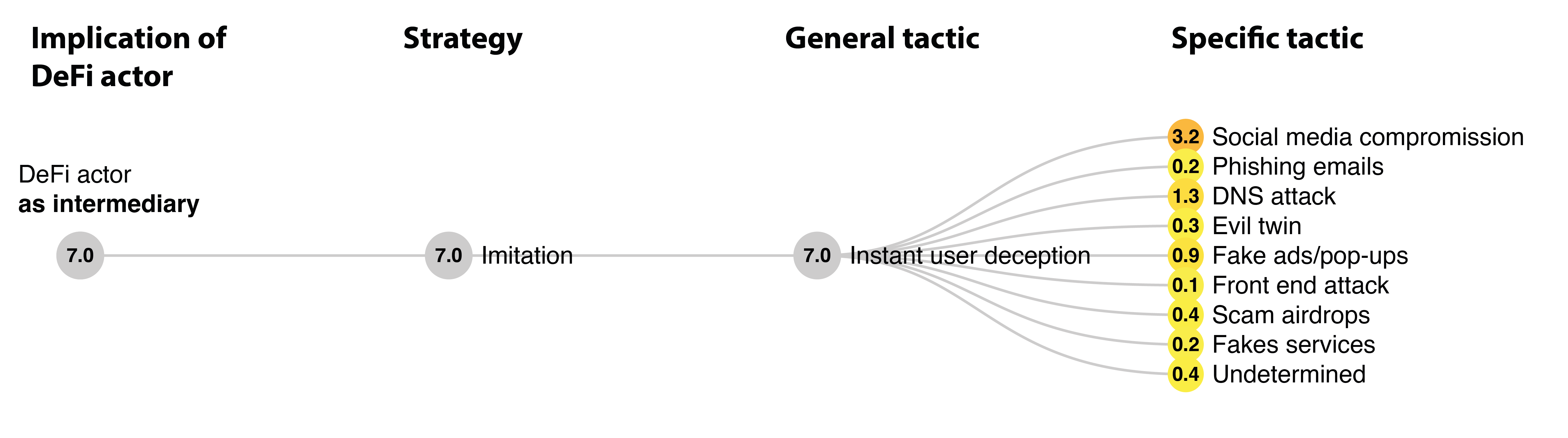}
	\caption{\textbf{DeFi Actors as Intermediary: Distribution of Strategies and Tactics}. Percentages are calculated based on the entire dataset. Our taxonomy shows that for the third scenario (DeFi actor as intermediary), the malicious source relies on an imitation strategy to ultimately defraud users. Further desegregating this strategy into general and specific tactics then provides additional insight on how the event can be conducted.}
	\label{fig:tax_actor_vector}
\end{figure}

The third implication was when DeFi actors were used to reach users (7\%). In such situations, DeFi actors served as intermediaries between the attacker and the targets; for instance, they might be imitated in fraudulent phishing schemes. Although DeFi actors could also suffer from such an attack, the ultimate target of the attack remained the users. The only strategy uncovered was imitation, where a malicious actor impersonated DeFi actors online to defraud users. The sole general tactic was user deception as the goal was always to trick users into believing that the information provided through deception was valid. Also, note that the specific tactics, presented in Figure~\ref{fig:tax_actor_vector}, either used the DeFi actor's compromised infrastructure or a  newly built infrastructure that resembled a legitimate DeFi actor's one to reach users.

Note that the low prevalence of DeFi actors as intermediaries in this dataset contradicts other studies focusing on phishing attacks, which highlight that such attacks are prevalent due to their simplicity and numerous opportunities~\cite{andryukhin_phishing_2019, phillips_tracing_2020} . This discrepancy arises because this dataset relied entirely on events reported by aggregators, as explained earlier, which might lack visibility into fraudulent platforms and impersonation schemes. Many of these schemes might also go unreported in the news media, leading to limited online visibility. It is likely that the dataset created for this study under represents the prevalence of such crime events. Therefore, such events are excluded from our subsequent DeFi technical stack mapping and stolen amount analysis. Future studies could enrich our dataset by adding such crime events and further comparing the three categories.

\subsection*{Technical impact assessment}

\begin{figure}
	\includegraphics[width=\textwidth]{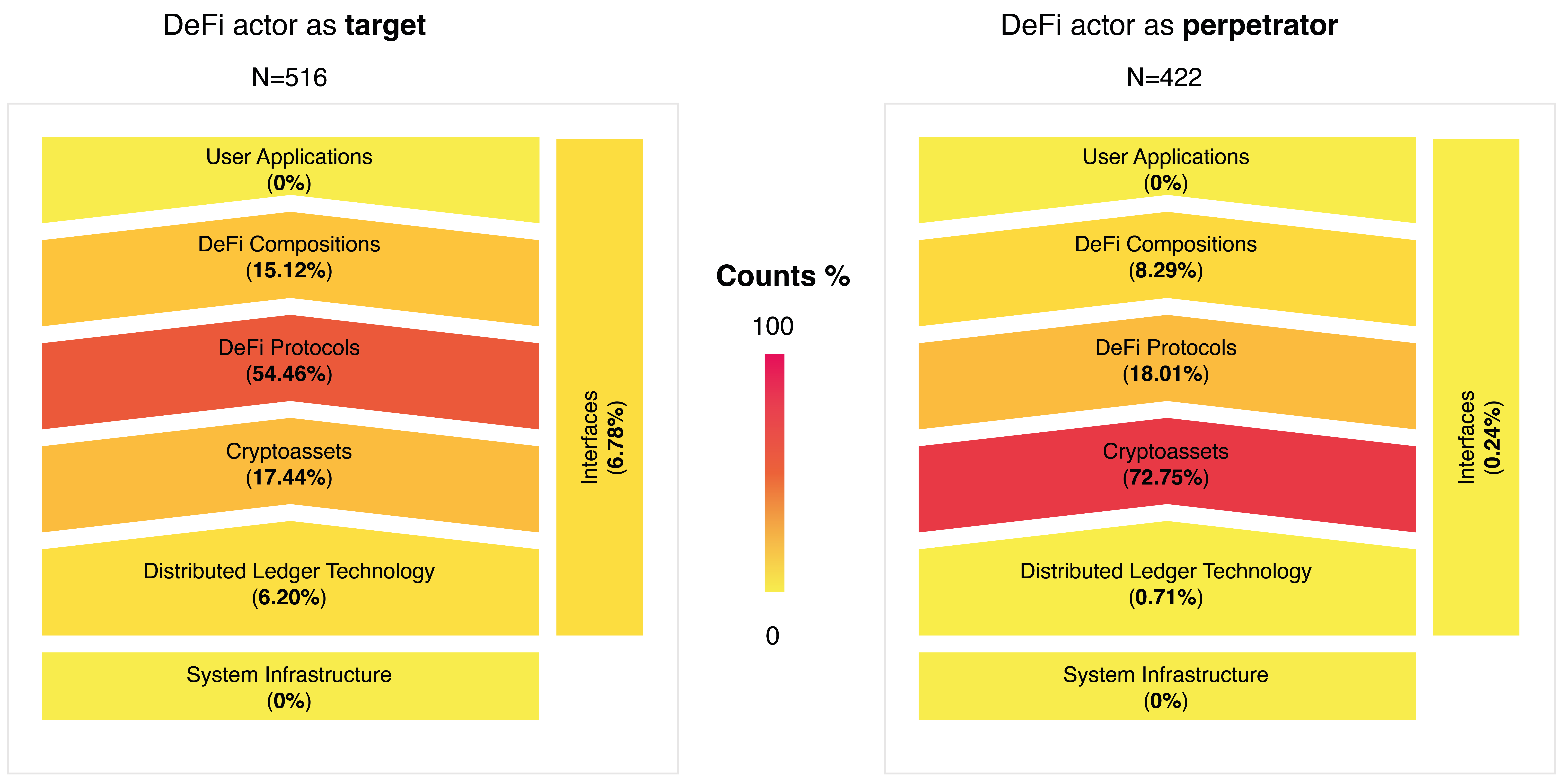}
 	\caption{\textbf{Affected technical layers}. This figure displays the technical layers of the Augmented DeFi Stack Reference Model (see Figure~\ref{fig:DeFi_stack_overview}) and illustrates the degree to which each layer has been impacted by DeFi crime events, according to the category of the DeFi actor involved. We differentiate between events where the DeFi actor was the victim and those where the actor played a perpetrator role.}
	\label{fig:defi_stack_impact}
\end{figure}

The second objective of this study is to determine which technical layers of the DeFi tech stack are most affected by crime events. Our analysis is based on the previously introduced Augmented DeFi Stack Reference (DSR) Model (refer to Figure~\ref{fig:DeFi_stack_overview}). We mapped each crime event to the most specific and relevant layer affected, focusing on the categories \emph{DeFi actor as target} and \emph{DeFi actor as perpetrator}. Starting from the 1036 crime events in the dataset, we also excluded actors that did not operate in one of the predefined categories presented in Table~\ref{tab:categories-actors} of section~\customref{subsec:method_obj2}{Obj~2}. Out of the remaining crime events that had a defined actor's category, we also removed all events categorized under \emph{DeFi actor as intermediary}, as while these events utilized the services of DeFi actors, they did not directly impact the actors themselves. Figure \ref{fig:defi_stack_impact} illustrates the relative distribution of these remaining 938 crime events across the technical layers and based on the implication of DeFi actors.

In the first scenario, titled \emph{DeFi actor as a target}, we analyzed a total of \NTarget/ crime events. Our analysis revealed that the majority of these attacks targeted the \emph{DeFi Protocols} layer, constituting 54.46\% of all events. The \emph{Cryptoassets} layer was the next most targeted, accounting for 17.44\%. The \emph{DeFi Compositions} layer was comparatively less affected, experiencing 15.12\% of the attacks. The \emph{Interfaces} to other systems were the subject of 6.78\% of the events, while the \emph{Distributed Ledger Technology} layer represented 6.20\%. Significantly, there were no documented attacks on the foundational \emph{System Infrastructure} or the \emph{User Application} layers. This is because we mapped the events according to the main area of operation of the DeFi actor involved, and no ``area of operation" is system infrastructure or user application, even if the event affected these layers. Indeed, these results denote which layers were attacked based on the category of the actor who bore the financial impact. Also layers external to the one corresponding to the actor's category can be affected or exploited during an event. As depicted in Figure~\ref{fig:implications}, third-party DeFi actors might be instrumentalized for an attack. For instance, a decentralized exchange (DEX) situated in the protocol layer could be targeted via oracle manipulation, inherently involving the interface layer. 

In the second scenario, titled \emph{DeFi actor as perpetrator}, we examined a total of 422 crime events. Our findings indicated that actions within the \emph{Cryptoassets} layer, notably custom token contracts, were the primary methods used to target other users, representing an astonishing 72.75\% of the events. Both the \emph{DeFi Protocols} and \emph{Protocol Compositions} layers were less frequently utilized, each comprising 18.01\% and 8.29\% of the events respectively. The \emph{Interfaces} to other systems and the foundational \emph{Distributed Ledger Technology} had minimal involvement of 0.24\% and 0.71\%, thus both playing a rather subordinate role in this scenario.

\subsection*{Size and Scope of Financial Damages}

The third objective of this study is to quantify the financial damages associated with these crimes. To begin, Table~\ref{tab:financial_damages_strategies} presents the sum of financial damages based on the position of actors in crime events, as well as the strategies used to steal money. To do so, the sample uses the crime events mapped onto the technical stack and keeps only the crime events that displayed a financial damage amount. As shown in Table~\ref{tab:financial_damages_strategies}, crime events targeting DeFi actors resulted in a loss of \AActorTargetFinD/, of which technical vulnerabilities accounted for  \ATargetTechnicalVulnerability/ of damages and  \ATargetHumanRisk/ to human risk. Note that 83\% of all financial damages related to crime events directly targeted DeFi actors. On the other hand, crime events where the DeFi actor was the perpetrator led to losses of \AActorPerpetratorFinD/ (the resulting 17\%). Within these, malicious use of contracts represented about half of the financial damages (\APerpetratorMaliciousUseofContract/), and market manipulation constituted the other half (\APerpetratorMarketManipulation/). In total, from 2017 to 2022, crime events where DeFi actors were targeted resulted in greater financial damages than those where DeFi actors themselves were themselves perpetrators.

\begin{table}
	\centering
	\begin{tabular*}{\columnwidth}{@{\extracolsep{\fill}}llcc}
		\toprule
	 	Scenarios & Strategies & Count & {Financial}  \\
	 	&  &  & {Damage}\\
		\midrule
		\textbf{DeFi actor as Target} & Technical vulnerability & \NTargetTechnicalVulnerability/&  \ATargetTechnicalVulnerability/ \\
		& Human risks & \NTargetHumanRisk/ & \ATargetHumanRisk/ \\ 
            & Undetermined & \NTargetUndetermined/ & \ATargetUndetermined/ \\ 
            \cline{3-4} 
		& & \NActorTargetFinD/ & \AActorTargetFinD/ \\	\\	
		\textbf{DeFi actor as Malicious} &  Malicious use of contracts &  \NPerpetratorMaliciousUseofContract/ &  \APerpetratorMaliciousUseofContract/\\
		&  Market manipulation &  \NPerpetratorMarketManipulation/& \APerpetratorMarketManipulation/ \\ \cline{3-4} 
		& & \NActorPerpetratorFinD/ &  \AActorPerpetratorFinD/ \\ 
		\bottomrule
	\end{tabular*}
	\caption{\textbf{Total Financial Damages in USD}. This table separates financial damages resulting from targeted DeFi actors from those resulting from DeFi perpetrators, along with financial damages specific to strategies. Overall, DeFi actors being targeted generated significantly more financial damages than DeFi actors acting maliciously.}
    \label{tab:financial_damages_strategies}

\end{table}

Examining the mean differences between these two categories presents an interesting narrative. Based on a Mann-Whitney U tests, crime events in which DeFi actors were targets ($\Tilde{x} = \$800,000$) resulted in significantly higher financial damages ($U=126 874, p=0.000, d=0.446$) per crime event than those in which DeFi actors were perpetrators ($\Tilde{x}=\$88,736$). 

\begin{figure}
	\includegraphics[width=0.95\columnwidth]{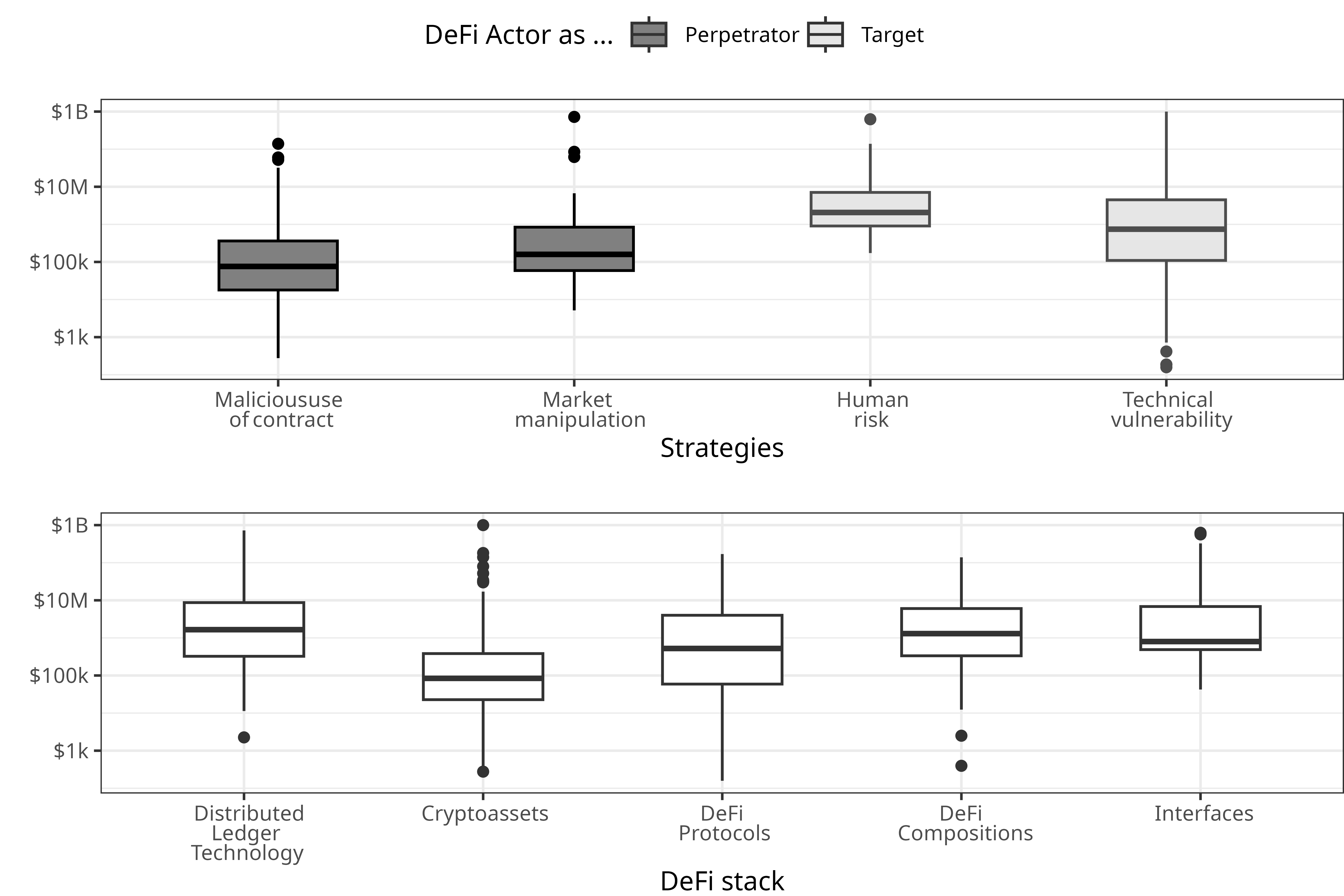}
	\caption{\textbf{Financial Damages Distributions for Strategies and Stack Layers} This figure visualizes, first, differences in financial damages between the strategies and implication of actors, and second, financial damages between the tech stack layers. In all cases, there exist significant differences between the implication of actors, the strategies used and the tech stack layer and their related financial damages.}
	\label{fig:boxplots}
\end{figure}

Figure~\ref{fig:boxplots} illustrates the distribution of financial damages based on the actors' implication and their related strategies. To determine whether the visible differences in the figure were statistically significant, we conducted a series of tests. Firstly, the Kruskall-Wallis test, which compared financial damages based on the main strategies used in the crime event, was significant ($H=139.04, p=0.000, \epsilon^2=0.17$).

Post-hoc test results suggested that exploiting human risks led to significantly more financial damages than all other strategies. Specifically, exploiting human risks ($\Tilde{x} = \$2,073,280$) led to significantly more financial damages ($p=0.000, d=0.78$) than malicious use of contracts ($\Tilde{x} = \$76,018$), significantly more damages ($p=0.027, d= 0.333$) than exploiting technical vulnerabilities ($\Tilde{x} = \$764,523$) and significantly more damages ($p=0.000, d= 0.638$) than market manipulation ($\Tilde{x} = \$159,136$).

Post-hoc test results also suggested that malicious use of contracts ($\Tilde{x} = \$76,018$) led to fewer financial damages compared to exploiting human risks (as mentioned above) but also led to significantly fewer financial damages ($p=0.000, d=0.45$) than exploiting technical vulnerabilities ($\Tilde{x} = \$744,000$). The remaining tests were non-significant.

Table~\ref{tab:financial_damages_stack} presents how the financial damages were distributed in the tech stack. The DeFi protocol layer faced the most financial damages with a total of \AStackP/, followed by the oracle/interface layer with \AStackINT/ of losses. Then, the cryptoasset layer faced \AStackCA/ of damages, followed by the DLT layer with \AStackDLT/ and, finally, the DeFi composition layer with \AStackCP/. However, note that the number of events happening on each layer differs, as discussed above.

\begin{table}
	\centering
	\begin{tabular*}{\columnwidth}{@{\extracolsep{\fill}}lcc}
		\toprule
		Category & Count & {Financial}  \\
		&  &  {Damage}\\
		\midrule
		DeFi Protocols ($DP$) & \NStackP/ & \AStackP/ \\
		Interface ($INT$) & \NStackINT/ & \AStackINT/ \\
		Cryptoassets ($CA$) & \NStackCA/ & \AStackCA/ \\
		Distributed Ledger Technology ($DLT$) & \NStackDLT/ & \AStackDLT/ \\
		DeFi Compositions ($DC$) & \NStackCP/ & \AStackCP/ \\
	\bottomrule
	\end{tabular*}
	\caption{\textbf{Total Financial Damages in USD in descending order.} This table shows the sum of financial damages that can be associated with the different stack layers. The Protocol and Interface layers faced the most extensive financial damages, even if the cryptoasset layer was involved in more crime events.}
	\label{tab:financial_damages_stack}
\end{table}

Figure~\ref{fig:boxplots} also shows the distribution of financial damages in the tech stack. Similarly to above, to assess whether the differences that can be seen in the figure are significant, we computed a series of statistical tests. The Kruskall-Wallis H test comparing financial damages according to where the crime event took place on the tech stack was also significant ($H=139.65, p=0.000, \epsilon^2=0.17$).

The results of the post-hoc tests showed that crimes targeting the Cryptoassets (CA) layers led to fewer financial damages than crimes targeting other layers. Indeed, with the post-hoc tests, we found that crime events targeting the cryptoasset (CA) layer ($\Tilde{x} = \$83,960$) led to significantly fewer damages ($p=0.000, d= 0.34$) than crimes targeting the Protocol (P) layer ($\Tilde{x} = \$522,807$), significantly fewer damages ($p=0.000, d= 0.66$) than crimes targeting the Interface layer ($\Tilde{x} = \$800,000$), significantly fewer damages  ($p=0.000, d= 0.62$) than crimes targeting the DeFi Composition (DC) layer ($\Tilde{x} = \$1,300,000$) and significantly fewer damages ($p=0.000, d= 0.54$) than crimes targeting the Distributed Ledger Technology (DLT) layer ($\Tilde{x} = \$1,655,628$). 

Crimes targeting the DeFi protocol layer (DP)  ($\Tilde{x} = \$522,807$) led to significantly fewer financial damages ($p=0.000, d= 0.27$) than crimes targeting the Interface (INT) layer  ($\Tilde{x} = \$800,000$) and significantly fewer damages ($p=0.000, d= 0.21$) than crimes targeting the DeFi Composition (DC) layer  ($\Tilde{x} = \$1,300,000$). The remaining tests were non-significant.

%% file: sections/discussion.tex
% !TeX root = ../main.tex

\section*{Discussion}

Previous studies have developed various categorizations to better understand crime events, from considering DeFi technologies as a facilitating tool for cybercrime versus a target of attack~\cite{reddy_analysing_2020} or differentiating between technical and market attacks~\cite{kris_oosthoek_flash_nodate, werner_sok_2022}. To move beyond these general categorizations, this study presents an innovative view on illicit activities taking place in the DeFi sector by combining the implications of DeFi actors in the crime event as well as the strategies and tactics used by malicious actors. The developed taxonomy, combined with interpretations on how it plays on the Augmented DeFi Stack Reference Model~\cite{auer_technology_2023}, as well as estimates on the total recorded financial damages, provides a comprehensive, informative and unique portrayal of the DeFi crime landscape. We present below the key study's takeaways, as well as how they are embedded with previous studies on the topic. 

\paragraph{CeFi: Important Financial Damages}
While significant efforts are directed towards research that aim at improving and securing DeFi services, our study highlights that the CeFi industry play a dominant role in financial damages related to cryptoassets. Indeed, the findings highlights that the entire cryptoasset industry suffered, from 2017 to 2022, a minimum loss of \$30 billion, with two thirds relating to CeFi and one third to DeFi. This finding is significant considering that the data collection focused on creating a comprehensive account of DeFi crime events, and while only 105 CeFi crime events were flagged during the collection, these events account for much more financial damages than the 1036 DeFi crime events This may be due to large-scale fraud such as Ponzi schemes that accounts for \NCeFiPonzi/ and embezzlement schemes that account for \NCeFiEmbezzlement/ of all the recorded stolen amounts. Given these differences in scale, potentially, the decentralized features provided by the DeFi technologies may, in the end, better protect users against such large-scale fraud. The DeFi industry may, moreover, be smaller in scale than the CeFi one. In the end, considering that, these days, most cryptoasset trading occurs off-chain via centralized exchanges~\cite{aramonte_defi_2021, auer_banking_2023}, subsequent studies should aim to further investigate the CeFi space, as these services might harbor greater vulnerabilities than anticipated.

\paragraph{Vulnerable DeFi Actors as Main Target for Profit-Driven Crime}
The results highlighted that \PActorTarget/\% of  crime events directly targeted DeFi actors and that these events caused 83\% (\AActorTargetFinD/ out of \$9.3B) of recorded financial damages in the DeFi sector. Moreover, crime events in which DeFi actors were targeted led to, on average, more financial damages than crime events in which DeFi actors were perpetrators. These results should not be underestimated: DeFi actors are the primary victims of crime taking place in the ecosystem. Similarly to our results, a private industry report~\cite{chainalysis_2022_2022} highlighted that the 10 most financially devastating attacks of 2020 were almost all carried out using price manipulation, manipulation of oracles, flash loans, and exploitation of vulnerabilities in smart contracts, which are all components of our technical vulnerability strategy. In our results, almost all attacks targeting DeFi actors leveraged a technical vulnerability, as opposed to a human risk. Moreover, note that a third of all crime events also pertained to vulnerabilities found in smart contracts. That DeFi projects are vulnerable and can be exploited is well-known\cite{ghaleb_towards_2022}. The alarming proportions of smart contract exploitations was also mentioned in various studies~\cite{wen_attacks_2021, ghaleb_towards_2022, qian_smart_2022}. One study also highlighted that inadequate authentication and authorization were major problems, as they were a frequent cause of vulnerability in smart contracts~\cite{chen_survey_2020}. This is also reflected in our findings, as access control flaws were the most prevalent specific tactic related to contract exploitation. This suggests a need for better access control mechanisms in contracts. 

Also, note that the second strategy related to DeFi actor as target, exploitation of human risks, yielded \ATargetHumanRisk/ in losses for \NTargetHumanRisk/  events. Hence, such strategy might be less prevalent, but it generates large financial losses per event. While human risks appear less prevalent in our dataset, it is also important to remember that 2.3\% of the events have an undetermined strategy. It is possible that some of those events occurred because of a human risk (such as employee deception) that was not disclosed, or has yet to be uncovered. Moreover, the study does not account for mixed strategies: some crime events may have had a human component that was not taken into account. Indeed, various social engineering techniques can be used in tandem with the tactics displayed in the taxonomy to steal money from DeFi actors and users. Further studies could investigate how, and to what extent, the crime events displayed in the dataset had a human component, beyond the obvious ones that were categorized under the exploiting human risk strategy.
In the end, the results suggest that DeFi actors should not solely focus on securing their smart contracts to mitigate external threats, but should also revisit and improve the precautions taken at the internal level.

\paragraph{DeFi actor as Perpetrator: Malicious uses of Contracts and Limited Financial Damages}

On the other hand, in about \PActorPerpetrator/\% of crime events, DeFi actors were themselves perpetrators, but the financial damages amounted to only 17\% (\AActorPerpetratorFinD/ out of \$9.3B) of recorded financial damages. These results differ from other studies that reported higher instances and higher impacts of malicious DeFi projects ~\cite{xia_trade_2021, la_morgia_pump_2020, mazorra_not_2022, mansourifar_hybrid_2020}. This may be because these other studies leveraged machine learning technologies to scan contracts and find malicious ones. However, a contract that may contain malicious code does not mean that such code was used. Also, what makes a contract malicious can sometimes be interpreted differently. For example, a liquidity function can be seen as a potential backdoor, but may also be considered a necessary function to control a token's liquidity. Also, note that an industry report stated that users lost about USD 2.8B when rug pulls were at their peak in 2021~\cite{chainalysis_2022_2022}. Such number largely differs from our findings of \AActorPerpetratorFinD/. However, the report does not provide any information on the methods used to aggregate the data, and also includes CeFi actors in the analysis, considering smart contract rug pulls and CeFi exit scam both as rug pulls. Such finding stresses the need for scientific and rigorous research on the topic. In this study, we recorded known events in which users lost money following their investments in the DeFi sector. These crime events are smaller in scale, as the results show that crime events in which the DeFi actors were perpetrators yielded fewer financial damages than crime events in which DeFi actors were legitimate and, subsequently, targeted. 

Also, the study's findings highlight that malicious uses of contracts, the main strategy identified, yielded fewer financial damages than other strategies. This may be because tools and techniques have been developed to detect them as many academic studies have investigated the topic~\cite{la_morgia_pump_2020,mazorra_not_2022,xia_trade_2021, mansourifar_hybrid_2020}. This is inspiring: there should be efforts to continuously develop such tools that identify contracts with functions that can be used for malicious purposes. Such tools should be disseminated to the general public. Still, if malicious activity is easily detected but not adequately reported, users are still at risk. This reinforces the relevance of developing mediums to efficiently broadcast such information to the community.    

Also, note that market manipulation, the second strategy related to DeFi actors as perpetrator, yielded \APerpetratorMarketManipulation/ in losses for only \NPerpetratorMarketManipulation/ events. Hence, the individual financial impact of such crime events is high. Similarly to above, these results therefore highlight that users should be careful in deciding what to invest in, as fraudulent investment projects can result in large losses.

\paragraph{Main Impacted Layers: DeFi protocol, Cryptoasset and Interface layers} 

The results of this study also shed light on the most vulnerable layers of the tech stack, considering the actor's implication in the crime events. In terms of prevalence, when actors are targeted, the most impacted DeFi tech stack layer is the DeFi protocols and when DeFi actors are perpetrators, the cryptoasset layer. The latter can be explained by the fact that our crime event dataset includes mostly rug pull scams. These crime events involved fungible tokens, which are a specific type of cryptoasset. The impact on the DeFi protocols layer is also intuitive as this layer encompasses the core financial functions (e.g., token swaps, borrowing, lending, etc.) offered by DeFi services. In DeFi-specific crime events, these financial functions thus represent natural targets. In terms of financial damages, DeFi protocols have faced most damages, with a total loss of \$2.6B. The cryptoasset layer, on the other hand, faced \$1.8B of losses. Hence, to mitigate victimization, it might be essential to emphasize the vulnerabilities of these two layers.

Our results also showed that the Interface layer faced high financial damages in a relatively small number of events (36): \$2.5B. These results show partial similarities with~\cite{zhou_sok_2022}. These authors found that yield farming and bridges accounted for a large proportion of losses (44\%) related to the studied DeFi incidents. However, note that previous studies investigating blockchain security~\cite{chen_survey_2020, duan_multiple-layer_2022, li_survey_2022} relied on a slightly different stack. In these studies, the smart contract and application layers are separated~\cite{chen_survey_2020, duan_multiple-layer_2022, li_survey_2022}, while in our study, which builds on~\cite{auer_technology_2023}, smart contracts are inherent to the DLT application layer, which encompasses all applications like Cryptoassets, DeFi protocols, and DeFi compositions. Such studies point out that the application layer, including programmable currencies and finance, and the smart contract layer, were the most involved in crime events. These findings are similar to ours as we find that most actors impacted by crime events operated mainly either at the DeFi protocol or the Cryptoasset layer. Another study~\cite{zhou_sok_2022} leveraged a set of 181 crime events from similar sources to ours and observed that the smart contract layer (42\%), protocol layer (40\%), and auxiliary services layer (30\%) were the most common causes of incidents. That the smart contract and the protocol layer are most targeted aligns with our findings. However, this study~\cite{zhou_sok_2022}  shows a higher percentage at the auxiliary service layer (30\%), compared to our study which finds limited crime events at the Interface layer (7\%). Such discrepancy could be explained by the authors' decision to include CeFi businesses and wallets as auxiliary services on top of off-chain oracles and cross-chain bridges, while we discarded non-DeFi specific related events from our stack mapping. In addition, our mapping was based on the category of the involved DeFi actor rather than the cause of the incident, as oracle services can be utilized rather than targeted. It is also relevant to mention that while all these studies did not introduce a clear distinction between targeted DeFi actors and DeFi perpetrators, some do simultaneously include targeted and fraudulent DeFi projects in their studied crime events~\cite{duan_multiple-layer_2022, zhou_sok_2022, li_survey_2022, chen_survey_2020}

%% file: sections/conclusion.tex
% !TeX root = ../main.tex
\section*{Limits and Further Research}

This study is the first of its kind, providing a comprehensive overview of crimes targeting the DeFi industry. However, several limitations must be acknowledged. First, the crime event dataset developed in this study depends on external aggregators, which themselves rely on publicly reported events. As a result, numerous crime events may have occurred without public disclosure, or minor events may not have received enough attention to be captured by these aggregators. Additionally, the results likely underestimate crime events involving DeFi intermediaries, as these are often not recorded by aggregators. Such events frequently involve phishing activities targeting clients and are typically extensive in both size and scope~\cite{kamps_moon_2018,phillips_tracing_2020,xia_characterizing_2020}. Further research should build on the dataset created (which is available online) and include such types of crime events. Moreover, further research could focus on developing a comprehensive dataset of CeFi crime events.

The taxonomy created also focused on identifying the main strategy and tactic (both general and specific) behind each crime event. However, some crime events may involve multiple strategies and tactics. Further research could explore how strategies and tactics are used in conjunction with one another. Moreover, social engineering may be underestimated in this study, as in some cases, the use of social engineering techniques might remain undisclosed or could be uncovered later. Online news articles and aggregators tend to report events as early as possible to quickly inform the DeFi community, but thorough and time-consuming investigations are sometimes necessary to produce an accurate post-mortem analysis. In fact, while most technical vulnerabilities and malicious transactions leave digital traces on the blockchain, this is not the case for vulnerabilities associated with human behavior. These may be more difficult to uncover quickly. Further research could focus on studying the use of social engineering as a primary tactic or as a secondary one in the crime events uncovered in this study. Additionally, our data was collected at a specific point in time, but the available information on crime events is dynamic. Therefore, it would be relevant to update the database regularly to observe, track, and account for changes across strategies and tactics.

Finally, the financial damages reported are dependent on the method used. In this study, we relied on aggregators and, subsequently, news sources. Further research could assess the accuracy of the reported amounts by conducting blockchain investigations. Additionally, given the large fluctuations in cryptoasset prices, there may be significant differences in the reported amounts. For example, in the case of the Ronin Bridge, some media reported a loss of \$551M~\cite{largest_crypto_hacks}, while others reported \$625M~\cite{axie_infinity_ronin_exploit}. This discrepancy was due to a price peak in ETH shortly after the hack. This is one example of how price fluctuations complicate the estimation process. While this limitation is inevitable, we can only aim to minimize discrepancies as much as possible.

\section*{Conclusion}

This study first highlights that the entire cryptoasset industry suffered, from 2017 to 2022, a minimum loss of \$30 billion, with two thirds relating to centralized finance (CeFi) and one third to DeFi. Focusing on the latter, this study examines crime events taking place in the DeFi sector by mapping their common tactics and strategies as well as the level of involvement for DeFi actors. This approach sheds light on victimized parties, whether DeFi actors or users. It also provides information on the most common methods taken by malicious actors to initiate such events, offering insights that can be leveraged to better secure the ecosystem. The dataset developed for this study is available online~\footnote{https://zenodo.org/records/14047933} to support research reproducibility and further analysis.

Our findings have implications for DeFi stakeholders. Given that DeFi actors are primary targets of attacks, \emph{regulatory authorities}, such as financial market authorities, should develop best practices in cybersecurity, encouraging measures like smart contract audits, cybersecurity training, and background checks for employees. These authorities could also consider fostering collaboration between traditional finance experts and DeFi platforms to develop standards that address technical vulnerabilities without stifling innovation. Since DeFi actors can also serve as intermediaries or perpetrators, \emph{regulatory authorities} should enhance efforts to raise user awareness about the risks associated with investing in DeFi activities. This can be achieved through targeted educational campaigns and clear guidelines on identifying potential scams or fraudulent behavior. The results of this study can be used in such campaigns as estimates of the size and scope of predatory profit-driven crime in the DeFi industry. Finally, for \emph{end users} and \emph{investors}, our findings indicate that using DeFi services is risky and could result in a complete loss of invested funds. While attack vectors such as rug pulls may seem trivial, they are often difficult to detect, even for technically savvy users. Therefore, DeFi services should be approached with great caution.

%% file: sections/annex.tex
% !TeX root = ../main.tex

\clearpage

\section{Appendix}

\subsection{Crime Events Statistics}\label{subsection:crimeevents}

\begin{table}[htb]
	\centering
	\begin{tabular*}{\columnwidth}{@{\extracolsep{\fill}}lcccccc}
		\toprule
		& Min & Mean & Med & Std & Max & \textbf{Total} \\
		\midrule
		Damage & \$158 & \$29.9M & \$345k & \$207M & \$3.6B & \$29.5B \\  
		\bottomrule
	\end{tabular*}
	\caption{\textbf{Descriptive statistics on crime financial damages in USD}. This table shows basic statistics on the financial damages reported by the events in our dataset. Note that information on damages was only available for \NallDbHasFinDam/ out of \NallDb/ events.}
	\label{tab:descrip-financial-damages-fin}
\end{table}

\pagebreak
\input{tables/Category_table_edited}

\pagebreak

\subsection{General and Specific Tactics Definitions}\label{subsection:definitions}

\setlength\LTleft{-\oddsidemargin}
\setlength\LTright{-\rightmargin}

% .... \paperwidth

\newcolumntype{R}[1]{>{\RaggedRight}p{#1}}
\newcolumntype{L}[1]{>{\RaggedLeft}p{#1}}

\begin{longtable}{
		>{\arraybackslash}m{2.3cm}
		>{\arraybackslash}m{4cm}|
		>{\arraybackslash}m{2.5cm}
		>{\arraybackslash}m{5.5cm}
	}
	\textbf{General\-tactic~} & \textbf{Definition} & \textbf{Specific-\-tactic~} & \textbf{Definition}   \\ \hline
	
	\multirow{5}{=}[1em]{\textbf{Contract vulner\-ability}} \vfill & 
		\multirow{5}{=}[1em]{A vulner\-ability in a smart contract’s code is exploited for theft purpose} \vfill & 
			\textbf{Re\-entrancy} \vfill & \vfill
				A withdraw function is repeatedly called before the vulnerable contract updates its balance. \vfill \\
	&	& 	\textbf{Access control flaw} \vfill & 
				Insufficient permission checks allowing privileged terms or functions to be called.  \vfill \\ 
	&	&	\textbf{Logical bug/\-custom flaw} \vfill &
				Wrong ordering of a smart contract code lines, or design flaw in the logic of the code leading to unintended or unexpected behavior. \vfill \\
	& 	& 	\textbf{Rollback} \vfill &
				Defrauding a lottery game without paying the bet cost by rolling back the corresponding unsatisfied reversible transaction. \vfill \\
	& 	& 	\textbf{Random number} & 
				A malicious actor exploits the vulnerabilities in a procedure to generate random numbers. The random value is predicted by running a weak pseudo-random number generator (PRNG). \vfill \\
	&  	& 	\textbf{K-value verification vulner\-ability} \vfill & 
				K value is not verified properly \vfill \\ 
	\hline
	\multirow{3}{=}[1em]{\textbf{Hacked/\-exploited infra\-structure}} &
		\multirow{3}{=}[1em]{A vulner\-ability in the infrastructure of a DeFi actor’s platform is exploited and allows malicious activities to be carried out}
			& \textbf{Accessing private keys/\-data} \vfill & \vfill A malicious actor accesses sensitive information stored by a DeFi actor. \vfill \\
	&  	&	\textbf{BGP hi\-jack} \vfill & Internet traffic is rerouted by a malicious actor. \vfill \\
	& 	& 	\textbf{Ransom\-ware} \vfill & A DeFi actor is black\-mailed after a malicious actor accessed some of their sensitive information. \vfill \\ 
	\hline
	
	\multirow{2}{=}[1em]{
	\textbf{Inter\-connected actors flaw}} & 
		\multirow{2}{=}[1em]{A loophole in a financial service implicating multiple DeFi actors leads to a vulnerability that is exploited by a malicious actor} & 
			\textbf{Flash loan arbitrage} \vfill & 		
				\vfill Performing arbitrage or exploit poorly designed economic model with flash loan funded capital to maximize profit. \vfill
	\\
	& 	& 	\textbf{Oracle manipu\-lation} & 
		Profiting from an oracle routing incorrect price information to a DeFi actor’s smart contract, or an inefficient oracle price feed. \vfill \\ 
	\hline

	\multirow{3}{=}[1em]{\textbf{Transaction attack}} & 
		\multirow{3}{=}[1em]{Malicious actors’ profit from blockchain’s transaction order process to carry out malicious activities} & 
			\textbf{Replay} \vfill & 
				\vfill Intercepting and copying a user’s transaction data and replay it on another blockchain.	\vfill \\
	&  	& 	\textbf{Front\-running} \vfill & 
				Front-run a transaction on hold by leveraging gas fees importance. \vfill \\
	& 	& 	\textbf{Trans\-action congestion} \vfill & Sending an abnormal number of small transactions on a platform to mislead smart contracts. \vfill
	\\ 
	\hline
	
	\multirow{2}{=}[1em]{\textbf{Dec\-entralization issue}} & 
		\multirow{2}{=}[1em]{The governance or consensus mechanism of a DeFi actor is targeted to carry out malicious activities} & 
			\textbf{51\% attack} 
   \vfill & 
				A malicious user gains control of more than 50\% of the mining power in a blockchain \vfill \\
	& 	& 	\textbf{Vote manipu\-lation} \vfill & 
				Taking over a smart contract or reshaping the rules by initiating a proposal with the certainty of obtaining the required votes. \vfill \\ 
	\hline
	
	\multirow{4}{=}[2em]{\textbf{Internal theft}} & 
		\multirow{4}{=}[2em]{An insider commits theft by taking advantage of their strategic position in the DeFi actor’s organization to perform unauthorized operation for personal gains} & 
			\textbf{Un\-authorized use of private key} \vfill & 
				\vfill A team member accesses a DeFi actor’s private keys and performs unauthorized transactions for his personal benefits. \vfill \\
	&	& 	\textbf{Contract vulnerability exploit} \vfill & 
				A team member discovers and exploits a contract vulnerability, instead of disclosing or fixing it. \vfill \\ 
	&	& 	\textbf{Malicious code injection} \vfill & 
				A rogue developer injects a malicious code in a smart contract during its deployment or while its being upgraded to steal assets or drain the contract. \vfill  \\
	&	& 	\textbf{Backdoor} \vfill & 
				A team member inserts a backdoor during the development of a smart contract. \vfill \\ 
	\hline
	
	\multirow{2}{=}[1em]{\vfill \textbf{External factors} \vfill} \vspace{1.5cm}  & 
		\multirow{2}{=}[2em]{A malicious actor obtains a DeFi actor’s sensitive information that allows him to commit theft by taking advantage of a platform’s mistakes or by directly targeting employees} \vfill & 
			\textbf{Exploiting operational mistake} \vfill & 
				\vfill Team member compromises sensitive information by mishandling it or storing it ineffectively. \vfill \\
	&	&	\textbf{Deceiving \linebreak personnel} \vfill \vspace{2cm} & 
				Team member is deceived into granting funds or data access to a malicious external party. \vfill  \\ 
    
	\hline

	\multirow{1}{=}[4em]{\textbf{Un-\\determined}} \vfill &
	
	It is unclear if the event is rooted in technical vulnerabilities or human risks. A malicious party obtains private keys or data by uncertain means, either by deceiving personnel or exploiting the DeFi actor's infrastructure  & 
	\textbf{Accessing private keys/\-data} \vfill \vspace{4cm} & \vspace{5cm}
	\vfill \\

	\hline
	
	\pagebreak
	
	\multirow{4}{=}[2em]{\textbf{Rug pull scam}} & 
		\multirow{4}{=}[2em]{A project’s creator performs malicious actions with its project’s assets or inserts malicious terms/functions in the project’s code to perform an exit scam and defraud users} & 
			\textbf{Liquidity \linebreak removal} & 
				\vfill The creator removes its share of liquidity from the project’s liquidity pool for personal profit. \vfill \\
	&	&	\textbf{Selling \linebreak restrictions} \vfill & 
				The creator disables the transfer function, which restricts users into selling their assets. \vfill \\
	& 	& 	\textbf{Hidden mint function} \vfill & 
				The creator implements a hidden mint function that enables him to selfishly get additional tokens. \vfill \\
	&	&	\textbf{Pump and dump} \vfill & 
				The creator works on increasing the value of his project before selling all of his personal assets at the inflated price. \vfill \\ 
	\hline
	
	\multirow{3}{=}[3em]{\textbf{Mis\-appropriation of funds}} & 
		\multirow{3}{=}[3em]{A project’s creator uses investors’ funds for personal gain} & 
			\textbf{Ponzi scheme} \vfill & 
				\vfill Creators ensure steady personal profit with a constant flow of investors. Funds obtained from recent investors are routed as profit to earlier ones to maintain the illusion of a successful project. \vfill \\
	&	& 	\textbf{Embezzlement} \vfill & 
				Part of investors' funds are being used for different purposes than the DeFi actor’s legitimate business activities. \vfill \\
	&	& 	\textbf{Scam presale, initial DEX offering (IDO) and initial coin offering (ICO)} \vfill & 
				Users purchase assets prior to their release date or donate to raise funds for a startup protocol but the creator never delivers the project. \vfill \\ 
	\hline
	
	\multirow{8}{=}[1em]{\textbf{Instant user deception}} & 
		\multirow{8}{=}[1em]{Users are prompt to interact with content that seems to be displayed by legitimate DeFi actors. Ultimately, they are deceived by malicious actors, who gain access to their personal information or their assets.} & 
			\textbf{Social media compromission} \vfill & \vfill A DeFi actor’s social media account is hijacked and used to post malicious material that aims to defraud users. \vfill \\
	& 	& 	\textbf{Phishing emails} \vfill & 
				Phishing emails are sent to a DeFi actor’s users. \vfill \\
	& 	& 	\textbf{DNS attack} \vfill & 
				The website’s home page of a DeFi actor is altered to redirect users to phishing content. \vfill \\
	& 	& 	\textbf{Evil twin} \vfill & 
				Fake websites using techniques like typosquatting deceive users into thinking it is the legitimate DeFi actor’s website. \vfill \\
	& 	& 	\textbf{Fake \linebreak ads/pop-ups} \vfill & 
				Ads are purchased for phishing websites to be displayed amongst the first search results. \vfill \\
	& 	& 	\textbf{Scam airdrops} \vfill & 
				Phishing tokens are airdropped to users, who are redirected to a fake website when trying to redeem them. Their approval allows the malicious actor to take control of their wallet. \vfill \\
	& 	& 	\textbf{Front-end \linebreak attack} \vfill & 
				A vulnerability in the front end is used to upload malicious files on a server to deceive users. \vfill \\
	&	& 	\textbf{Fake services} \vfill & 
				A user invests on a platform, but the intended service is never delivered as the platform was invented to defraud \vfill \\ 
	\hline

	\caption{\textbf{General and specific tactics definitions}. For concision purposes, this table regroups and defines all possible general and specific tactics referenced in Fig~\ref{fig:tax_actor_target}, ~\ref{fig:tax_actor_malicious} and ~\ref{fig:tax_actor_vector}, the two most disaggregated categories of our taxonomy. The general tactic refers to the broad method used by malicious actors to steal assets, while the specific tactic specifies the possible  techniques that can be used to achieve the former.}\\

\end{longtable}

%% file: tables/Category_table_edited.tex
% latex table generated in R 4.4.1 by xtable 1.8-4 package
% Mon Dec 16 15:15:52 2024

\begin{table}[ht]
\centering
\resizebox{\textwidth}{!}{%
\begin{tabular}{rllllllll}
  
 & 
 Paper\_category & 
 2017 & 
 2018 & 
 2019 & 
 2020 & 
 2021 & 
 2022 & 
 ALL \\ 
  
1 & 
 ALL & 
 8~\par{}{\scriptsize(\$246.1M}) & 
 54~\par{}{\scriptsize(\$1.2B}) & 
 74~\par{}{\scriptsize(\$192.4M}) & 
 208~\par{}{\scriptsize(\$1.4B}) & 
 281~\par{}{\scriptsize(\$2.8B}) & 
 411~\par{}{\scriptsize(\$4B}) & 
 1036~\par{}{\scriptsize(\$10B}) \\ 
  2 & 
 Blockchain & 
 - & 
 8~\par{}{\scriptsize(\$23.3M}) & 
 4~\par{}{\scriptsize(\$8.1M}) & 
 11~\par{}{\scriptsize(\$769.5M}) & 
 5~\par{}{\scriptsize(\$22.6M}) & 
 9~\par{}{\scriptsize(\$655.6M}) & 
 37~\par{}{\scriptsize(\$1.5B}) \\ 
  3 & 
 Bridge & 
 - & 
 - & 
 - & 
 1~\par{}{\scriptsize(\$0}) & 
 11~\par{}{\scriptsize(\$667.9M}) & 
 18~\par{}{\scriptsize(\$1.8B}) & 
 30~\par{}{\scriptsize(\$2.5B}) \\ 
  4 & 
 Dapp & 
 - & 
 33~\par{}{\scriptsize(\$7.5M}) & 
 59~\par{}{\scriptsize(\$9.1M}) & 
 14~\par{}{\scriptsize(\$26.6M}) & 
 27~\par{}{\scriptsize(\$395.5M}) & 
 61~\par{}{\scriptsize(\$142.4M}) & 
 194~\par{}{\scriptsize(\$581.1M}) \\ 
  5 & 
 Derivatives & 
 - & 
 - & 
 1~\par{}{\scriptsize(\$0}) & 
 2~\par{}{\scriptsize(\$419k}) & 
 2~\par{}{\scriptsize(\$6.7M}) & 
 5~\par{}{\scriptsize(\$21.5M}) & 
 10~\par{}{\scriptsize(\$28.7M}) \\ 
  6 & 
 Exchange & 
 1~\par{}{\scriptsize(\$251k}) & 
 3~\par{}{\scriptsize(\$23.6M}) & 
 - & 
 19~\par{}{\scriptsize(\$265.7M}) & 
 46~\par{}{\scriptsize(\$529M}) & 
 47~\par{}{\scriptsize(\$406.4M}) & 
 116~\par{}{\scriptsize(\$1.2B}) \\ 
  7 & 
 FT & 
 2~\par{}{\scriptsize(\$33.5M}) & 
 5~\par{}{\scriptsize(\$1.1B}) & 
 3~\par{}{\scriptsize(\$5M}) & 
 116~\par{}{\scriptsize(\$34.3M}) & 
 88~\par{}{\scriptsize(\$115.3M}) & 
 159~\par{}{\scriptsize(\$407.9M}) & 
 373~\par{}{\scriptsize(\$1.7B}) \\ 
  8 & 
 Lending & 
 - & 
 - & 
 - & 
 13~\par{}{\scriptsize(\$180.5M}) & 
 19~\par{}{\scriptsize(\$376.8M}) & 
 25~\par{}{\scriptsize(\$191.8M}) & 
 57~\par{}{\scriptsize(\$749.1M}) \\ 
  9 & 
 NFT & 
 - & 
 - & 
 - & 
 2~\par{}{\scriptsize(\$8M}) & 
 6~\par{}{\scriptsize(\$5.7M}) & 
 37~\par{}{\scriptsize(\$48.1M}) & 
 45~\par{}{\scriptsize(\$61.8M}) \\ 
  10 & 
 Oracle & 
 - & 
 - & 
 - & 
 1~\par{}{\scriptsize(\$270.7k}) & 
 3~\par{}{\scriptsize(\$1.8M}) & 
 3~\par{}{\scriptsize(\$1.6M}) & 
 7~\par{}{\scriptsize(\$3.7M}) \\ 
  11 & 
 Other & 
 5~\par{}{\scriptsize(\$212.4M}) & 
 5~\par{}{\scriptsize(\$51.5M}) & 
 7~\par{}{\scriptsize(\$170.1M}) & 
 13~\par{}{\scriptsize(\$60M}) & 
 7~\par{}{\scriptsize(\$1.8M}) & 
 13~\par{}{\scriptsize(\$49.2M}) & 
 50~\par{}{\scriptsize(\$545.2M}) \\ 
  12 & 
 Staking & 
 - & 
 - & 
 - & 
 1~\par{}{\scriptsize(\$2.5k}) & 
 10~\par{}{\scriptsize(\$110.1M}) & 
 11~\par{}{\scriptsize(\$70.8M}) & 
 22~\par{}{\scriptsize(\$180.9M}) \\ 
  13 & 
 Yield & 
 - & 
 - & 
 - & 
 15~\par{}{\scriptsize(\$86.1M}) & 
 57~\par{}{\scriptsize(\$582M}) & 
 23~\par{}{\scriptsize(\$221.5M}) & 
 95~\par{}{\scriptsize(\$889.6M}) \\ 

\end{tabular}
}
\end{table}

%% file: main.bbl
\begin{thebibliography}{10}

\bibitem{largest_crypto_hacks}
Largest crypto hacks - crypto \& blockchain security, 2023.
\newblock Accessed: 2024-10-18.

\bibitem{a_deshmukh_blockchain_2022}
{A. Deshmukh}, {N. Sreenath}, {A. K. Tyagi}, and {U. V. Eswara Abhichandan}.
\newblock Blockchain {Enabled} {Cyber} {Security}: {A} {Comprehensive}
  {Survey}.
\newblock In {\em 2022 {International} {Conference} on {Computer}
  {Communication} and {Informatics} ({ICCCI})}, pages 1--6, Coimbatore, India,
  January 2022. IEEE.
\newblock Journal Abbreviation: 2022 International Conference on Computer
  Communication and Informatics (ICCCI).

\bibitem{alshater_initial_2023}
MM~Alshater, M~Joshipura, R~El~Khoury, and N~Nasrallah.
\newblock Initial {Coin} {Offerings}: a {Hybrid} {Empirical} {Review}.
\newblock {\em Small Business Economics}, pages 1--18, 2023.

\bibitem{andryukhin_phishing_2019}
A.A. Andryukhin.
\newblock Phishing {Attacks} and {Preventions} in {Blockchain} {Based}
  {Projects}.
\newblock In {\em 2019 {International} {Conference} on {Engineering}
  {Technologies} and {Computer} {Science} ({EnT})}, pages 15--19, Moscow,
  Russia, March 2019. IEEE.

\bibitem{anita_blockchain_2019}
N~Anita. and M~Vijayalakshmi.
\newblock Blockchain {Security} {Attack}: {A} {Brief} {Survey}.
\newblock In {\em 2019 10th {International} {Conference} on {Computing},
  {Communication} and {Networking} {Technologies} ({ICCCNT})}, pages 1--6,
  Kanpur, India, July 2019. IEEE.

\bibitem{aramonte_defi_2021}
Sirio Aramonte, Wenqian Huang, and Andreas Schrimpf.
\newblock {DeFi} risks and the decentralisation illusion.
\newblock {\em BIS Quarterly Review}, December 2021.

\bibitem{atzei_survey_2017}
Nicola Atzei, Massimo Bartoletti, and Tiziana Cimoli.
\newblock A {Survey} of {Attacks} on {Ethereum} {Smart} {Contracts} ({SoK}).
\newblock In Matteo Maffei and Mark Ryan, editors, {\em Principles of
  {Security} and {Trust}}, Lecture {Notes} in {Computer} {Science}, pages
  164--186, Berlin, Heidelberg, 2017. Springer.

\bibitem{auer_banking_2023}
Raphael Auer, Marc Farag, Ulf Lewrick, Lovrenc Orazem, and Markus Zoss.
\newblock Banking in the {Shadow} of {Bitcoin}? {The} {Institutional}
  {Adoption} of {Cryptocurrencies}, 2023.

\bibitem{auer_technology_2023}
Raphael Auer, Bernhard Haslhofer, Stefan Kitzler, Pietro Saggese, and Friedhelm
  Victor.
\newblock The technology of decentralized finance ({DeFi}).
\newblock {\em Digital Finance}, August 2023.

\bibitem{barone_cryptocurrency_2019}
R~Barone and D~Masciandaro.
\newblock Cryptocurrency or usury? {Crime} and alternative money laundering
  techniques.
\newblock {\em European Journal of Law and Economics}, 47(2):233--254, April
  2019.

\bibitem{bartoletti_dissecting_2020}
Massimo Bartoletti, Salvatore Carta, Tiziana Cimoli, and Roberto Saia.
\newblock Dissecting {Ponzi} schemes on {Ethereum}: {Identification}, analysis,
  and impact.
\newblock {\em Future Generation Computer Systems}, 102:259--277, January 2020.

\bibitem{noauthor_south_2021}
Bloomberg.
\newblock South african brothers vanish, and so does \$3.6 billion in bitcoin.
\newblock June 2021.

\bibitem{caldarelli_blockchain_2021}
Giulio Caldarelli and Joshua Ellul.
\newblock The {Blockchain} {Oracle} {Problem} in {Decentralized}
  {Finance}—{A} {Multivocal} {Approach}.
\newblock {\em Applied Sciences}, 11(16):7572, January 2021.
\newblock Number: 16 Publisher: Multidisciplinary Digital Publishing Institute.

\bibitem{cao_survey_2022}
XL~Cao, JH~Zhang, XC~Wu, and B~Liu.
\newblock A survey on security in consensus and smart contracts.
\newblock {\em Peer-to-Peer Networking and Applications}, 15(2):1008--1028,
  March 2022.

\bibitem{cao_flashot_2021}
Yixin Cao, Chuanwei Zou, and Xianfeng Cheng.
\newblock Flashot: {A} {Snapshot} of {Flash} {Loan} {Attack} on {DeFi}
  {Ecosystem}, January 2021.
\newblock arXiv:2102.00626 [q-fin].

\bibitem{chainalysis_2022_2022}
Chainalysis.
\newblock The 2022 {Geography} of {Cryptocurrency} {Report}.
\newblock Technical report, Chainalysis, October 2022.

\bibitem{chen_survey_2020}
HS~Chen, M~Pendleton, L~Njilla, and SH~Xu.
\newblock A {Survey} on {Ethereum} {Systems} {Security}: {Vulnerabilities},
  {Attacks}, and {Defenses}.
\newblock {\em ACM Computing Surveys}, 53(3), June 2020.

\bibitem{noauthor_top_nodate}
Coinmarketcap.
\newblock Top cryptocurrency exchanges ranked by volume.
\newblock \url{https://coinmarketcap.com/rankings/exchanges/}, Accessed
  2023-02-15.
\newblock A list of cryptocurrency exchanges ranked by volume, including
  Binance, Coinbase Pro, Huobi, and more.

\bibitem{noauthor_check_nodate}
{CoinMarketCap}.
\newblock Historical market cap snapshots of cryptocurrencies.
\newblock \url{https://coinmarketcap.com/historical/}, Accessed 2023-10-05.
\newblock Historical market cap snapshots of cryptocurrencies, starting in
  April 2013. See all time high crypto prices from 2017 and 2018. Bitcoin.
  Ethereum. More.

\bibitem{defi_announcing_2023}
DeFi.
\newblock Announcing {The} {World}’s {First} {DeFi} {REKT} {Database}, May
  2023.

\bibitem{noauthor_fi_nodate}
De.Fi.
\newblock De.fi - defi investing \& yield farming platform.
\newblock \url{https://de.fi/rekt-database}, Accessed 2023-09-01.

\bibitem{noauthor_defillama_nodate}
DefiLlama.
\newblock Defillama: A defi tvl aggregator.
\newblock \url{https://defillama.com/}, Accessed 2023-09-01.
\newblock DefiLlama is a DeFi TVL aggregator committed to providing accurate
  data without ads or sponsored content, as well as transparency.

\bibitem{duan_multiple-layer_2022}
Li~Duan, Yangyang Sun, Kejia Zhang, and Yong Ding.
\newblock Multiple-{Layer} {Security} {Threats} on the {Ethereum} {Blockchain}
  and {Their} {Countermeasures}.
\newblock {\em Security and Communication Networks}, 2022:e5307697, February
  2022.
\newblock Publisher: Hindawi.

\bibitem{dunn_estimation_1959}
Olive~Jean Dunn.
\newblock Estimation of the {Medians} for {Dependent} {Variables}.
\newblock {\em The Annals of Mathematical Statistics}, 30(1):192--197, 1959.
\newblock Publisher: Institute of Mathematical Statistics.

\bibitem{dunn_multiple_1961-1}
Olive~Jean Dunn.
\newblock Multiple {Comparisons} among {Means}.
\newblock {\em Journal of the American Statistical Association},
  56(293):52--64, March 1961.
\newblock Publisher: Taylor \& Francis \_eprint:
  https://www.tandfonline.com/doi/pdf/10.1080/01621459.1961.10482090.

\bibitem{ghaleb_towards_2022}
Asem Ghaleb.
\newblock Towards {Effective} {Static} {Analysis} {Approaches} for {Security}
  {Vulnerabilities} in {Smart} {Contracts}.
\newblock In {\em 37th {IEEE}/{ACM} {International} {Conference} on {Automated}
  {Software} {Engineering}}, pages 1--5, Rochester MI USA, October 2022. ACM.

\bibitem{gridley_significant_2023}
Jared Gridley and Oshani Seneviratne.
\newblock Significant {Digits}: {Using} {Large}-{Scale} {Blockchain} {Data} to
  {Predict} {Fraudulent} {Addresses}, January 2023.
\newblock arXiv:2301.01809 [cs].

\bibitem{axie_infinity_ronin_exploit}
Sandali Handagama.
\newblock Axie infinity's ronin network suffers \$625m exploit, 2022.
\newblock Accessed: 2024-10-18.

\bibitem{hendrickson_cash_2022}
JR~Hendrickson and WJ~Luther.
\newblock Cash, crime, and cryptocurrencies.
\newblock {\em Quarterly Review of Economics and Finance}, 85:200--207, August
  2022.

\bibitem{huang_geography_2018}
Winifred Huang, Michele Meoli, and Silvio Vismara.
\newblock The {Geography} of {Initial} {Coin} {Offerings}, July 2018.

\bibitem{kamps_moon_2018}
Josh Kamps and Bennett Kleinberg.
\newblock To the moon: defining and detecting cryptocurrency pump-and-dumps.
\newblock {\em Crime Science}, 7(1):1--1, November 2018.

\bibitem{kleinberg_cryptocurrencies_2022}
Arianna Kleinberg, Bennett Trozze, and Josh Kamps.
\newblock Cryptocurrencies: Boons and curses for fraud prevention.
\newblock In {\em A Fresh Look at Fraud}. Routledge, 2022.
\newblock Num Pages: 28.

\bibitem{kris_oosthoek_flash_nodate}
{Kris Oosthoek}.
\newblock Flash {Crash} for {Cash}: {Cyber} {Threats} in {Decentralized}
  {Finance}.

\bibitem{kyngas_inductive_2020}
Helvi Kyngäs.
\newblock Inductive {Content} {Analysis}.
\newblock In Helvi Kyngäs, Kristina Mikkonen, and Maria Kääriäinen,
  editors, {\em The {Application} of {Content} {Analysis} in {Nursing}
  {Science} {Research}}, pages 13--21. Springer International Publishing, Cham,
  2020.

\bibitem{la_morgia_pump_2020}
M~La~Morgia, A~Mei, F~Sassi, J~Stefa, and {IEEE}.
\newblock Pump and {Dumps} in the {Bitcoin} {Era}: {Real} {Time} {Detection} of
  {Cryptocurrency} {Market} {Manipulations}.
\newblock In {\em 2020 29th {International} {Conference} on {Computer}
  {Communications} and {Networks} ({ICCCN})}, Honolulu, HI, USA, 2020. IEEE.

\bibitem{li_security_2022}
Wenkai Li, Jiuyang Bu, Xiaoqi Li, and Xianyi Chen.
\newblock Security {Analysis} of {DeFi}: {Vulnerabilities}, {Attacks} and
  {Advances}.
\newblock In {\em 2022 {IEEE} {International} {Conference} on {Blockchain}
  ({Blockchain})}, pages 488--493, Espoo, Finland, August 2022. IEEE.

\bibitem{li_survey_2022}
Wenkai Li, Jiuyang Bu, Xiaoqi Li, Hongli Peng, Yuanzheng Niu, and Yuqing Zhang.
\newblock A {Survey} of {DeFi} {Security}: {Challenges} and {Opportunities},
  2022.
\newblock Publisher: arXiv.

\bibitem{liang_data-driven_2021}
Yuzhi Liang, Weijing Wu, Kai Lei, and Feiyang Wang.
\newblock Data-driven {Smart} {Ponzi} {Scheme} {Detection}, August 2021.
\newblock arXiv:2108.09305 [cs].

\bibitem{mackenzie_criminology_2022}
Simon Mackenzie.
\newblock Criminology {Towards} the {Metaverse}: {Cryptocurrency} {Scams},
  {Grey} {Economy} and the {Technosocial}.
\newblock {\em British Journal of Criminology}, 62(6):1537--1552, November
  2022.

\bibitem{mansourifar_hybrid_2020}
Hadi Mansourifar, Lin Chen, and Weidong Shi.
\newblock Hybrid {Cryptocurrency} {Pump} and {Dump} {Detection}, March 2020.
\newblock arXiv:2003.06551 [cs].

\bibitem{matakovic_crypto-assets_2022}
Ivana~Cunjak Mataković.
\newblock Crypto-{Assets} {Illicit} {Activities}: {Theoretical} {Approach} with
  {Empirical} {Review}.
\newblock {\em International e-Journal of Criminal Sciences}, (17):1--30,
  January 2022.

\bibitem{mazorra_not_2022}
Bruno Mazorra, Victor Adan, and Vanesa Daza.
\newblock Do {Not} {Rug} on {Me}: {Leveraging} {Machine} {Learning}
  {Techniques} for {Automated} {Scam} {Detection}.
\newblock {\em Mathematics}, 10(6):949, January 2022.
\newblock Number: 6 Publisher: Multidisciplinary Digital Publishing Institute.

\bibitem{meissel2024using}
Kane Meissel and Esther~S Yao.
\newblock Using cliff’s delta as a non-parametric effect size measure: an
  accessible web app and r tutorial.
\newblock {\em Practical Assessment, Research, and Evaluation}, 29(1), 2024.

\bibitem{naylor_towards_2003}
R.~T. Naylor.
\newblock Towards a {General} {Theory} of {Profit}‐{Driven} {Crimes}.
\newblock {\em The British Journal of Criminology}, 43(1):81--101, January
  2003.

\bibitem{naylor_predators_2003}
R.T. Naylor.
\newblock Predators, {Parasites}, or {Free}-{Market} {Pioneers}: {Reflections}
  on the {Nature} and {Analysis} of {Profit}-{Driven} {Crime}.
\newblock In Margaret~E. Beare, editor, {\em Critical {Reflections} on
  {Transnational} {Organized} {Crime}, {Money} {Laundering}, and {Corruption}},
  pages 35--54. University of Toronto Press, 2003.

\bibitem{nghiem_detecting_2021}
Huy Nghiem, Goran Muric, Fred Morstatter, and Emilio Ferrara.
\newblock Detecting cryptocurrency pump-and-dump frauds using market and social
  signals.
\newblock {\em Expert Systems with Applications}, 182:115284, November 2021.

\bibitem{nolasco_braaten_convenience_2021}
Claire Nolasco~Braaten and Michael~S. Vaughn.
\newblock Convenience {Theory} of {Cryptocurrency} {Crime}: {A} {Content}
  {Analysis} of {U}.{S}. {Federal} {Court} {Decisions}.
\newblock {\em Deviant Behavior}, 42(8):958--978, August 2021.
\newblock Publisher: Routledge \_eprint:
  https://doi.org/10.1080/01639625.2019.1706706.

\bibitem{noauthor_cryptosec_2021}
Cryptosec (now ChainSec).
\newblock Cryptosec – crypto, blockchain, web3, cbdc, digital assets
  cybersecurity and investigations.
\newblock \url{https://chainsec.io/defi-hacks/}, Accessed 2024-09-01.

\bibitem{ostertagova_methodology_2014}
Eva Ostertagová, Oskar Ostertag, and Jozef Kováč.
\newblock Methodology and {Application} of the {Kruskal}-{Wallis} {Test}.
\newblock {\em Applied Mechanics and Materials}, 611:115--120, 2014.
\newblock Publisher: Trans Tech Publications Ltd.

\bibitem{paquet-clouston_ransomware_2019}
Masarah Paquet-Clouston, Bernhard Haslhofer, and Benoît Dupont.
\newblock Ransomware payments in the {Bitcoin} ecosystem.
\newblock {\em Journal of Cybersecurity}, 5(1):1--11, January 2019.

\bibitem{paquet-clouston_spams_2019}
Masarah Paquet-Clouston, Matteo Romiti, Bernhard Haslhofer, and Thomas Charvat.
\newblock Spams meet {Cryptocurrencies}: {Sextortion} in the {Bitcoin}
  {Ecosystem}.
\newblock In {\em Proceedings of the 1st {ACM} {Conference} on {Advances} in
  {Financial} {Technologies}}, {AFT} '19, pages 76--88, New York, NY, USA,
  October 2019. Association for Computing Machinery.

\bibitem{phillips_tracing_2020}
Ross Phillips and Heidi Wilder.
\newblock Tracing {Cryptocurrency} {Scams}: {Clustering} {Replicated}
  {Advance}-{Fee} and {Phishing} {Websites}.
\newblock In {\em 2020 {IEEE} {International} {Conference} on {Blockchain} and
  {Cryptocurrency} ({ICBC})}, pages 1--8, May 2020.

\bibitem{pieroni_smarter_2018}
Alessandra Pieroni, Noemi Scarpato, Luca Di~Nunzio, Francesca Fallucchi, and
  Mario Raso.
\newblock Smarter {City}: {Smart} {Energy} {Grid} based on {Blockchain}
  {Technology}.
\newblock {\em International Journal on Advanced Science, Engineering and
  Information Technology}, 8(1):298, February 2018.

\bibitem{puggioni_crypto_2022}
Valerio Puggioni.
\newblock Crypto rug pulls: {What} is a rug pull in crypto and 6 ways to spot
  it, February 2022.

\bibitem{qian_smart_2022}
Peng Qian, Zhenguang Liu, Qinming He, Butian Huang, Duanzheng Tian, and Xun
  Wang.
\newblock Smart {Contract} {Vulnerability} {Detection} {Technique}: {A}
  {Survey}, September 2022.

\bibitem{qin_cefi_2021}
Kaihua Qin, Liyi Zhou, Yaroslav Afonin, Ludovico Lazzaretti, and Arthur
  Gervais.
\newblock {CeFi} vs. {DeFi} -- {Comparing} {Centralized} to {Decentralized}
  {Finance}, June 2021.
\newblock arXiv:2106.08157 [cs, q-fin].

\bibitem{qin_attacking_2021}
KH~Qin, LY~Zhou, B~Livshits, and A~Gervais.
\newblock Attacking the {DeFi} {Ecosystem} with {Flash} {Loans} for {Fun} and
  {Profit}.
\newblock pages 3--32, Berlin, Heidelberg, 2021. Springer Berlin Heidelberg.

\bibitem{ramos_great_2021}
S~Ramos, F~Pianese, T~Leach, and E~Oliveras.
\newblock A great disturbance in the crypto: {Understanding} cryptocurrency
  returns under attacks.
\newblock {\em Blockchain-Research and Applications}, 2(3), September 2021.

\bibitem{reddy_analysing_2020}
E~Reddy.
\newblock Analysing the {Investigation} and {Prosecution} of {Cryptocurrency}
  {Crime} as {Provided} for by the {South} {African} {Cybercrimes} {Bill}.
\newblock {\em Statute Law Review}, 41(2):226--239, June 2020.

\bibitem{reddy_cryptocurrency_2018}
Eveshnie Reddy and Anthony Minnaar.
\newblock Cryptocurrency : a tool and target for cybercrime.
\newblock {\em Southern African Journal of Criminology}, 31(3):71--92, December
  2018.

\bibitem{s_shukla_addressing_2022}
{S. Shukla}, {I. Gupta}, and {K. Naresh}.
\newblock Addressing {Security} {Issues} and {Future} {Prospects} of {Web} 3.0.
\newblock In {\em 2022 2nd {Asian} {Conference} on {Innovation} in {Technology}
  ({ASIANCON})}, pages 1--7, Ravet, India, August 2022. IEEE.
\newblock Journal Abbreviation: 2022 2nd Asian Conference on Innovation in
  Technology (ASIANCON).

\bibitem{shier_statistics_2004}
Rosie Shier.
\newblock Statistics: 2.3 {The} {Mann}-{Whitney} {U} {Test}, 2004.

\bibitem{noauthor_slowmist_nodate}
Slowmist.
\newblock Slowmist: A blockchain security firm.
\newblock \url{https://slowmist.medium.com}, Accessed 2023-06-02.

\bibitem{noauthor_slowmist_nodate-1}
SlowMist.
\newblock Slowmist hacked - slowmist zone.
\newblock \url{https://hacked.slowmist.io/en/}, Accessed 2023-09-01.
\newblock A platform providing comprehensive information on various hacking
  incidents in the blockchain space.

\bibitem{tomczak2014need}
Maciej Tomczak and Ewa Tomczak.
\newblock The need to report effect size estimates revisited. an overview of
  some recommended measures of effect size.
\newblock 2014.

\bibitem{wang_blockeye_2021}
Bin Wang, Han Liu, Chao Liu, Zhiqiang Yang, Qian Ren, Huixuan Zheng, and Hong
  Lei.
\newblock {BLOCKEYE}: {Hunting} for {DeFi} {Attacks} on {Blockchain}.
\newblock In {\em 2021 {IEEE}/{ACM} 43rd {International} {Conference} on
  {Software} {Engineering}: {Companion} {Proceedings} ({ICSE}-{Companion})}.
  IEEE, May 2021.

\bibitem{wang_impact_2022}
Ye~Wang, Patrick Zuest, Yaxing Yao, Zhicong Lu, and Roger Wattenhofer.
\newblock Impact and {User} {Perception} of {Sandwich} {Attacks} in the {DeFi}
  {Ecosystem}.
\newblock In {\em Proceedings of the 2022 {CHI} {Conference} on {Human}
  {Factors} in {Computing} {Systems}}, {CHI} '22, pages 1--15, New York, NY,
  USA, April 2022. Association for Computing Machinery.

\bibitem{wen_attacks_2021}
Yujuan Wen, Fengyuan Lu, Yufei Liu, and Xinli Huang.
\newblock Attacks and countermeasures on blockchains: {A} survey from layering
  perspective.
\newblock {\em Computer Networks}, 191:107978, May 2021.

\bibitem{werner_sok_2022}
Sam~M. Werner, Daniel Perez, Lewis Gudgeon, Ariah Klages-Mundt, Dominik Harz,
  and William~J. Knottenbelt.
\newblock {SoK}: {Decentralized} {Finance} ({DeFi}), September 2022.
\newblock arXiv:2101.08778 [cs, econ, q-fin].

\bibitem{xia_trade_2021}
Pengcheng Xia, Haoyu wang, Bingyu Gao, Weihang Su, Zhou Yu, Xiapu Luo, Chao
  Zhang, Xusheng Xiao, and Guoai Xu.
\newblock Trade or {Trick}? {Detecting} and {Characterizing} {Scam} {Tokens} on
  {Uniswap} {Decentralized} {Exchange}, November 2021.
\newblock arXiv:2109.00229 [cs].

\bibitem{xia_characterizing_2020}
Pengcheng Xia, Haoyu Wang, Bowen Zhang, Ru~Ji, Bingyu Gao, Lei Wu, Xiapu Luo,
  and Guoai Xu.
\newblock Characterizing cryptocurrency exchange scams.
\newblock {\em Computers \& Security}, 98:101993, November 2020.

\bibitem{zetzsche_decentralized_2020}
Dirk~A Zetzsche, Douglas~W Arner, and Ross~P Buckley.
\newblock Decentralized {Finance}.
\newblock {\em Journal of Financial Regulation}, 6(2):172--203, September 2020.
\newblock Publisher: Oxford University Press (OUP).

\bibitem{zhou_dependability_2021}
CC~Zhou, LD~Xing, and QS~Liu.
\newblock Dependability {Analysis} of {Bitcoin} subject to {Eclipse} {Attacks}.
\newblock {\em International Journal of Mathematical Engineering and Management
  Sciences}, 6(2):469--479, April 2021.

\bibitem{zhou_sok_2022}
Liyi Zhou, Xihan Xiong, Jens Ernstberger, Stefanos Chaliasos, Zhipeng Wang,
  Ye~Wang, Kaihua Qin, Roger Wattenhofer, Dawn Song, and Arthur Gervais.
\newblock {SoK}: {Decentralized} {Finance} ({DeFi}) {Attacks}, September 2022.
\newblock arXiv:2208.13035 [cs].

\end{thebibliography}
